\documentclass[a4paper,fleqn,usenatbib,useAMS]{mnras}

\usepackage{etoolbox}
\makeatletter
\patchcmd\@combinedblfloats{\box\@outputbox}{\unvbox\@outputbox}{}
\makeatother
\usepackage{graphicx}	
\usepackage{amsmath}	
\usepackage{amssymb}	
\usepackage{multicol}   
\usepackage{bm}		    
\usepackage{subfig}
\usepackage{pdflscape}	

\usepackage[T1]{fontenc}
\usepackage{ae,aecompl}

\usepackage{newtxtext,newtxmath}

\title[Outside the Lyman-break box]{Outside the Lyman-break box: detecting Lyman continuum emitters at $3.5<z<5.1$ with CLAUDS}

\author[U. Me\v{s}tri\'{c} et al.]{
Uro\v{s} Me\v{s}tri\'{c},$^{1,2}$\thanks{E-mail: umestric@swin.edu.au or umestric@yandex.ru},
E. V. Ryan-Weber$^{1,2}$,
J. Cooke$^{1,2}$,
R. Bassett$^{1,2}$,
M. Sawicki$^{3,4,5}$\thanks{Canada Research Chair}, \newauthor
A. L. Faisst,$^{6}$
K. Kakiichi,$^{7}$
A. K. Inoue,$^{8,9}$
M. Rafelski,$^{10,11}$
L. J. Prichard,$^{10}$ \newauthor
S. Arnouts$^{12}$
T. Moutard$^{3}$
J. Coupon$^{13}$
A. Golob$^{3}$
S. Gwyn$^{14}$
\\
$^{1}$Centre for Astrophysics and Supercomputing, Swinburne University of 
Technology, Hawthorn, VIC 3122, Australia\\
$^{2}$ARC Centre of Excellence for All Sky Astrophysics in 3 Dimensions (ASTRO 3D), Australia\\
$^{3}$Institute for Computational Astrophysics \& Department of Astronomy and Physics, Saint Mary's University, Halifax, Canada\\\
$^{4}$Department of Astronomy and Physics, Saint Mary's University, 923 Robie Street, Halifax, Nova Scotia, B3H 3C3, Canada\\
$^{5}$Herzberg Astronomy and Astrophysics, National Research Council of Canada, 5071 West Saanich Rd., Victoria, BC V9E 2E7, Canada\\
$^{6}$IPAC, California Institute of Technology 1200 E California Boulevard, Pasadena, CA 91125, USA\\
$^{7}$Department of Physics and Astronomy, University College London, London, WC1E 6BT, UK\\
$^{8}$Department of Physics, School of Advanced Science and Engineering, Waseda University, 3-4-1 Okubo, Shinjuku, 169-8555 Tokyo, Japan\\
$^{9}$Research Institute for Science and Engineering, Waseda University, 3-4-1 Okubo, Shinjuku, 169-8555 Tokyo, Japan\\
$^{10}$Space Telescope Science Institute, 3700 San Martin Drive, Baltimore, MD 21218, USA\\
$^{11}$Department of Physics \& Astronomy, Johns Hopkins University, Baltimore, MD 21218, USA\\
$^{12}$Aix Marseille Universit\'e, CNRS, Laboratoire d'Astro\-phy\-sique de Marseille, UMR 7326, F-13388,  Marseille, France\\
$^{13}$Astronomy Department, University of Geneva, Chemin d'Ecogia 16, CH-1290 Versoix, Switzerland\\
$^{14}$NRC-Herzberg, 5071 West Saanich Road, Victoria, British Columbia, V9E 2E7, Canada\\
}

\pubyear{2019}

\begin{document}
\label{firstpage}
\pagerange{\pageref{firstpage}--\pageref{lastpage}}
\maketitle

\begin{abstract}
Identifying non-contaminated sample of high-redshift galaxies with escaping Lyman continuum (LyC) flux is important for understanding the sources and evolution of cosmic reionization.  
We present CLAUDS $u$-band photometry of the COSMOS field to probe LyC radiation from spectroscopically confirmed galaxies at $z\geq3.5$ and outside the standard Lyman-break galaxy colour selection expectations.  
Complementary to the CLAUDS data, we use Subaru multi-filter photometry, \textit {Hubble Space Telescope} ($HST$) multi-filter imaging, and the spectroscopic surveys D10K, VUDS and 3D-HST.
We present a sample of Lyman continuum galaxy (LCG) candidates in the redshift range $3.5\lesssim z\lesssim5.1$.
Here, we introduce 5 LCG candidates, where two are flagged quality 1 and three quality 2.
The estimated $f_{\rm esc}^{\rm abs}$ for quality 1 candidates are in the range $\sim5\% - 73\%$ and $\sim30\% - 93\%$.
These estimates are based on our derived parameters from individual galaxies as inputs to a range of BPASS models as well as mean intergalactic medium (IGM) and maximal intergalactic and circumgalactic media (IGM+CGM) transmission.
We conclude that our search for LCGs is most likely biased to lines of sight with low HI densities or free from Lyman limit systems.
Our two best LCG candidates have EW (Ly$\alpha)\leq50$\AA\ and we find no correlation or anti-correlation between EW (Ly$\alpha$), $f_{\rm esc}^{\rm abs}$, and $R_{\rm obs}$, the ratio of ionizing to non-ionizing observed flux in the measured passbands.
Stacking candidates without solid LyC detections ($S/N<3$) results in an estimated $f_{\rm esc}^{\rm abs}$ from galaxies not greater than $1\%$.

\end{abstract} 




\section{Introduction}\label{1}

The Epoch of Reionization (EoR) is a critical period where structures in the Universe such as stars, galaxies, quasars and active galactic nuclei (QSO/AGN), started to form and evolve.  
The EoR can be considered a transition period from a neutral and opaque Universe towards the mostly transparent and ionized Universe that we observe today.  
This stage of the Universe's history took place between $6\lesssim z<15$, where $z\sim6$ is estimated from observations of the Gunn-Peterson trough in the spectra of QSOs \citep{Fan2002,FAN2006, Becker2015, Eilers2018, Bosman2018}.  
Similar results are obtained by searching for a drop in the fraction of the Ly$\alpha$ emitting galaxies, which predicts the end point of reionization in the range between $5.7<z<7$ \citep{Kashikawa2006,Ouchi2010,Ota2010, Faisst2014, Konno2018, Mason2018}.  
There are many indirect observational constraints that point to $z\sim 15$ as the most likely beginning of the EoR \citep{Greig2017, Bowman2018, PlanckCollaboration2018}.

Understanding the nature of the sources that emit Lyman continuum radiation (LyC) and reionized the Universe is one of the most persistent questions in modern astronomy.  LyC is UV radiation with energy $E\geq13.6$eV or $\lambda\leq912$ \AA\ that is able to ionize hydrogen.  
The first LyC photons are believed to have been emitted by massive objects like metal-free Population III stars \citep{Bromm2002,WYITHE2007,Ahn2012, Susa2014}.  
In later stages of the EoR, LyC radiation is thought to be primarily produced by O and B stars in young star-forming galaxies as well as AGN \citep{MADAU1999,MADAU2015}.  
An additional complexity is that the contribution to the ultraviolet background (UVB) from different sources is not constant, rather it changes as different populations of the objects evolve \citep{Wyithe2011,BECKER2013,Kakiichi2018}.
Although great progress has been made in the last twenty years, the question of which sources are responsible for reionization of the intergalactic medium (IGM) remains open.
 
Clarifying the relative contribution of the various sources of ionizing radiation, particularly the relative roles of star-forming galaxies and AGN, is still under debate \citep{Becker2015, MADAU2015}.
Various studies supported by observations indicate that the population of AGN are not sufficient to ionize the IGM at $z>3$ \citep{Haardt1996,Cowie2009,Fontanot2012,Grissom2014,Trebitsch2018, Kulkarni2019} and sufficiently contribute to the observed global UVB, as the contribution of AGN to the UVB peaks at $z\sim2$ \citep{COWIE2008}.  
However, some recent research points toward low luminosity AGN at high redshifts as a possible main driver of reionization at its early stages \citep{Giallongo2012,Giallongo2015,Grazian2018}.

Although AGNs may contribute to the UVB, currently there is wide acceptance that the major producers of LyC radiation are young star forming galaxies.
Therefore galaxies are considered the most likely sources responsible for driving the reionization of the Universe.  
Due to the fact that LyC radiation is in the far-UV part of the spectrum, we are limited to observing galaxies at roughly $3\lesssim z\lesssim4.5$ using ground-based telescopes. 
At $z>3$ the redshifted LyC flux falls in the optical part of the spectrum and can be observed from the ground.  
The $z\lesssim4.5$ limit comes from the fact that number density of systems containing neutral hydrogen rapidly increases at $z>4$ \citep{SARGENT1989,INOUE2008} and the chances of detecting LyC flux decreases to below 20\% \citep{INOUE2008}.

In the last 20 years, many observational (spectroscopic and photometric) efforts have been made to directly detect LyC flux from $z\sim$ 2.5 -- 4.5 galaxies to measure their escape fraction of ionizing flux ($f_{\rm esc}$) and to define the population of galaxies contributing to reionization \citep{LEITHERER1995,STEIDEL2001,Inoue2005,Iwata2009,Nestor2011,Vanzella2012,Grazian2012,Siana2015,Vasei2016,Rutkowski2017,Marchi2017,Marchi2018,Steidel2018, Fletcher2019,Bassett2019}. 
Until now, only a few have been successful in producing spectroscopic confirmations: Ion2 \citep{Vanzella2015,deBARROS2016,Vanzella2016}, Q1549-C25 \citep{Shapley2016}, Ion3 \citep{Vanzella2018}, as well recent ones \citep{Nakajima2019, Steidel2018}.
It is crucial to our understanding of the EoR and structure formation to identify a larger sample; however from previous results it can be concluded that developing efficient selection criteria for LyC emitting galaxies at $3<z<4$ remains a challenging task.

To date, galaxies examined for escaping LyC photons have been mostly selected using a variation of the Lyman break technique introduced by \cite{Steidel1996} or through narrow band selection of the Ly$\alpha$ emitting galaxies \citep[LAEs: e.g][]{Cowie1998,Iwata2009,Nestor2011}.
However it has been recognised that with Lyman break galaxy (LBG) selection, a significant number of high-redshift galaxies can escape the selection criteria \citep{STEIDEL1999,LeFEVRE2005}, as the technique is developed to be efficient but not comprehensive.  
The LBG colour criteria enclose an efficient selection region in colour-colour space where $z\sim3-4$ galaxies reside based on the assumption of zero escaping LyC flux.  
\cite{COOKE2014} applied various levels of LyC flux to LBG composite spectra and found that there is a notably large fraction ($\sim32\%$) of $z\sim3-4$ star forming galaxies that reside outside the standard LBG colour selection region.  Their colours are consistent with those of spectroscopically confirmed galaxies of \cite{LeFEVRE2005} randomly chosen in a 'blind' magnitude-limited survey (i.e., no colour selection) and galaxies selected via deep medium-band infrared photometry from the ZFOURGE survey \citep{Straatman2016}, complemented by $\sim$30 band photometry in the COSMOS, CDFS, and UDS legacy fields.
The high-redshift galaxies that fall outside the LBG selection box are estimated to have medium to high $f_{\rm esc}$ values and these newly-identified galaxies has been termed Lyman Continuum Galaxies \citep[LCGs;][]{COOKE2014}.

Complicating all selection methods for galaxies emitting LyC flux is the possibility of foreground contamination from low redshift objects. 
The probability of contamination by foreground objects increases at higher redshift. 
Due to this kind of contamination, non-ionizing emission from a foreground galaxy can easily be mistaken for LyC radiation.  
Research on the estimated probability of such contamination indicates that a non-negligible fraction ($\sim$7--13\%) of LyC candidates are contaminated by foreground galaxies  \citep{Siana2007,Vanzella2010,Mostardi2013,COOKE2014}.  
There are two ways to check high-redshift LyC candidates for contamination.
The first is to use deep spectroscopy of LCG candidates to search for low redshift galaxy features and the second is using high-spatial resolution space-based imaging i.e \textit{Hubble Space Telescope} (HST), since the contamination rate is proportional to the point spread function \citep{Vanzella2012,Siana2015}.

Another way to overcome the difficulties of studying high redshift star-forming galaxies is to study their proxies in the local Universe.  
However, complications while studying low redshift star-burst galaxies arise from the fact that they are mostly opaque to the ionizing radiation that is generated in them \citep{Grimes2009}.  
Promising LCG counterparts at $z<1$ are the Green Pea galaxies (GPs), introduced by \cite{Cardamone2009}, because they are compact objects with intense star-formation rates.  
The high [OIII]/[OII] ratios, high densities and possible presence of shocks found in some GPs indicate that they may be leaking ionizing photons into the IGM \citep{Jaskot2013,Nakajima2014,Fiasst2016}.  
Research by \cite{Izotov2016,Izotov2018A} reports the detection of escaping LyC radiation from four compact star-forming galaxies (SFGs)\footnote[1]{Compact SFGs include GPs and luminous compact galaxies. 
Galaxies in the $z\sim0.0-0.6$ range with H$\beta$ line luminosity L(H$\beta)\geq10^{40.5}erg\:s^{-1}$ are named Luminous Compact Galaxies.  
General characteristics of the compact SFGs are strong emission lines in the optical part of the spectrum that are coming from HII regions, produced by ionizing radiation from O-stars \citep{Izotov2011}.} with $f_{\rm esc}\sim$ 6--13\%.  
More recently, the same group reported the highest $f_{\rm esc}=46\pm2\%$ detection to date in low redshift compact SFGs \citep{Izotov2018}.  
These recent observational results indicate that all compact SFGs with reported LyC detections in \cite{Izotov2016,Izotov2018} share the same properties.

Detailed spectral analysis reveals that for all compact star forming LyC leakers the equivalent width (EW) of Ly$\alpha$ increases with increasing LyC escape fraction.  
Ly$\alpha$ lines in LyC leakers are also found to be double peaked. 
A decrease in the peak separation between the red and blue peaks is also found to correlate with an increase in $f_{\rm esc}^{\rm abs}$  \cite{Verhamme2015,Verhamme2017,Kakiichi2019,Kimm2019} and on average Ly$\alpha$ escape fraction correlates with LyC escape fraction, indicating the link between the two escape processes.  
All of these results appear promising in terms of finding reliable selection criteria for high-redshift counterparts of LCGs.  
But first, all of these correlations need to be statistically verified from larger samples of the LyC leaking galaxies at redshifts beyond $z\sim3$.

These galaxies at $z\sim 3-5$ are recognized as lower redshift counterparts of the galaxies responsible for reionization of the Universe for which it is impossible to directly detect ionizing LyC radiation.
This is why it is crucial to our understanding of the EoR that large samples of LCGs are identified to measure ionizing LyC flux directly and to provide a sample for calibrating indirect indicators (colour, Ly$\alpha$ line properties, [OIII]/[OII] ratio, etc.) that will point to the leakage of ionizing LyC radiation from $z>6$ galaxies into IGM. 
However, currently developing efficient selection criteria for LCGs at $z\sim3-5$ remains a challenging task.

In this work, we present results from our search for Lyman continuum ionizing radiation from $3.5<z<5.5$ LCGs using CLAUDS photometry with the aim to test the hypothesis that these galaxies reside outside the standard LBG selection box or LBG selection expectations.
In Section \ref{2} we explain the sample selection, and in Section \ref{3} we describe the method for LyC flux measurements. 
We present sub-sample analyses in Section \ref{4}, discuss our results in Section \ref{5}, and summarize our findings in Section \ref{6}.

\section{Data and Sample selection}\label{2}

In order to detect LyC radiation and estimate $f_{\rm esc}$ from galaxies at $z\geq3.5$ to study their contribution to the global budget of ionizing photons, we select and analyse candidates from the well-characterized COSMOS field \citep{Scoville2007}. 
The motivations for choosing the COSMOS field are the following: 

\begin{itemize}
  \item The availability of spectroscopic surveys: the DEIMOS 10k spectroscopic survey \citep[hereafter D10K;][]{Hasinger2018A}, VIMOS Ultra Deep Survey \citep[VUDS;][]{LeFevre2015,Tasca2017} and as well as 3D-HST grism spectroscopy \citep{Brammer2012,Momcheva2016},

  \item The space-based high-resolution imaging coverage of the whole field in at least one $HST$ filter, ACS F814W \citep{Koekemoer2007, Scoville2007/2, Massey2010}, and over large areas in F125W, F160W, F140W, F606W, F336W and F435W,

  \item Access to the CFHT Large Area $u$-band deep survey \citep[CLAUDS;][] {Sawicki2019},

  \item The ultra-deep photometry from Hyper Suprime-Cam (HSC) Subaru Strategic Program public data release 1 \citep[PDR1;][]{Aihara2018,2Aihara2018}.
\end{itemize}

We note that to ensure the CLAUDS $u$-band detections are attributed correctly to pure LyC flux, it is necessary to have spectroscopic redshifts.  
The spectroscopic redshifts of our sample were taken from the literature; we acknowledge that getting these from the literature introduces biases in galaxy selection methods.  
The goal of this paper is to start with a conservatively-selected spectroscopic parent sample and inspect the $u$-band images for LyC flux.  
The details of the sample selection are described below.

\subsection{Photometric and spectroscopic data}

The photometric data we use to probe LyC flux in our sample are provided by the CLAUDS survey whose astrometry is matched to the Subaru HSC DR1 data. 
The minimum depth of the CLAUDS imaging in the COSMOS field is $u_{\rm AB}\sim27.2$ AB \citep[$5\sigma$ in $2^{\prime\prime}$ diameter aperture;][]{Sawicki2019} and it is comparable in depth with the HSC-Deep program.

The MegaCam \citep{Boulade2003} wide field optical imager mounted on the CFHT telescope is used during CLAUDS observations.  
The quantum efficiency of MegaCam ranges from $35\%$ to $60\%$ at $3500\mathring{A}-4000\mathring{A}$, which makes MegaCam the most sensitive wide-field imager in this blue-optical wavelength range on any current telescope.
The data from the new CLAUDS $u$-band filter used in this work in comparison with older $u^*$-filter has two important advantages for our research; the new $u$- filter probes bluer wavelengths and has a sharp cut-off at $\sim$4000\AA\ and no red leak, whereas the older $u^*$ filter probes redder wavelengths and has red leak at $\sim5000$\AA.

\begin{figure}
  \centering
  \includegraphics[width=0.55\textwidth]{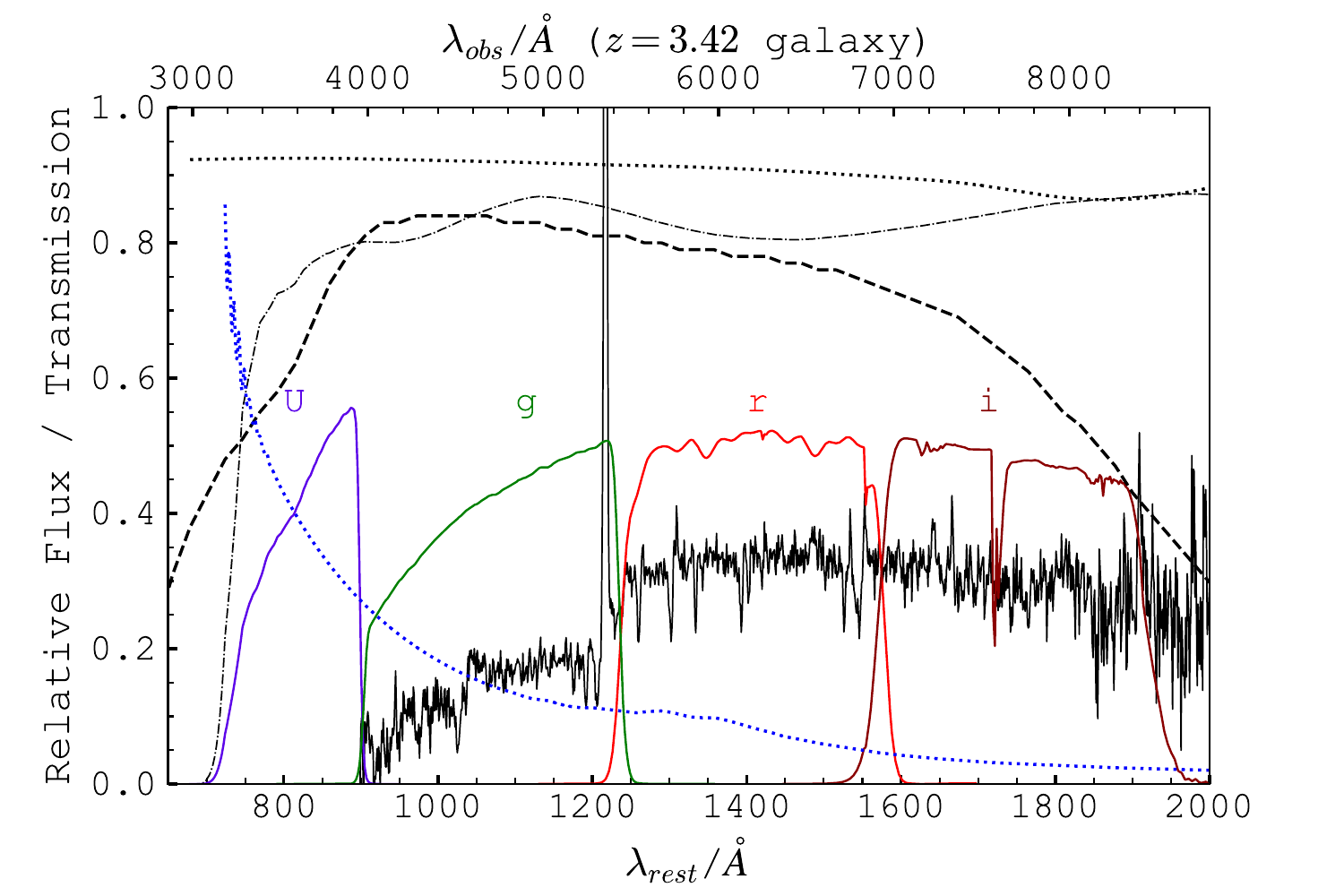}
  \caption{The composite LBG spectrum from \citep{Shapley2003} redshifted at $z=3.42$ (black) and transmission curves of the filters used in this work: CLAUDS $u$ (blue curve) and Subaru HSC $g$, $r$ and $i$ (green, orange and red curves, respectively).  
  The composite LBG spectrum is overlaid and illustrates that the $u$ filter probe the LyC flux and $g$, $r$, $i$ filters probe the regions of non-ionizing UV flux.  
  The sharp drop in sensitivity of the $u$-band filter at longer wavelengths ensures that objects at $z\sim3.42$ (where transmission of the filter is less than 0.5\%) are free from flux contamination by non-ionizing flux longward of 912\AA\ (i.e., the Ly$\alpha$ forest part).  
  The shape of the transmission curve of the $u$ filter is derived taking into account transmission from the CFHT primary mirror (black dotted line), MegaPrime optics (dot-dashed line) and CCD quantum efficiency (dashed line).  
  The blue dotted line represents median atmospheric extinction in mag/airmass units at the Mauna Kea site \citep{Buton2012}.}
  \label{fig1}
\end{figure}

Our aim is to test the hypothesis that LCGs at $z\geq3.5$ with measurable LyC emission, have colours that reside in locations outside of the LBG colour-selection region as proposed by \cite{COOKE2014}, see the Parent Sample subsection \ref{2.2} for more details. 
The lowest secure redshift at which we can select our candidates is $z=3.42$ as defined by the shape of the Megacam new $u$-band filter transmission curves.
Above this redshift, only the LyC part of the spectrum is probed by the $u$-band, with no chance of contamination by the Ly$\alpha$ forest part ($912 \mathring{A}< \lambda <1216 \mathring{A}$), even if the spectroscopic redshift is overestimated by $\Delta z\sim0.1$.  
Figure \ref{fig1} shows the transmission curves of the $u$-band filter overlaid on an LBG spectrum in rest frame wavelengths, alongside the transmission curves for the Subaru HSC $r$ and $i$ filters.  
In addition, the transmission curves of the CFHT primary mirror, MegaPrime optics, CCD quantum efficiency, and the atmospheric extinction for the Mauna Kea site are also shown. 

The Subaru HSC is a wide-field camera mounted on the Subaru 8.2 meter telescope (Hawaii, Mauna Kea site). 
The goal of the HSC survey is to observe high latitude fields like COSMOS in multiple photometric broad band ($g$,$r$,$i$,$z$,$y$) and narrow band filters \citep{2Aihara2018}.  
The filter depths based on PDR1 are 27.4, 27.3, 27, 26.4 and 25.6 mag for $g$, $r$, $i$, $z$ and $y$ respectively, \citep[$5\sigma$ depth for a point source;][]{2Aihara2018}).  Table \ref{tab1} summarises the photometric filter properties used in this work.

\begin{table}
\centering
\caption{Photometric filter characteristics}
\label{tab1}
\begin{tabular}{ccccc}
\hline
\hline
filter & \begin{tabular}[c]{@{}c@{}}$\lambda_{\rm eff}$\\ (\AA)\end{tabular} & \begin{tabular}[c]{@{}c@{}}FWHM\\ (\AA)\end{tabular} & \begin{tabular}[c]{@{}c@{}}$5\sigma$ depth\\ (mag)\end{tabular} & survey     \\ \hline
$u$ & 3538 &  860 & 27.2 & CLAUDS $^1$\\
$g$ & 4754 & 1395 & 27.4 & Subaru HSC $^{2}$ \\
$r$ & 6175 & 1503  & 27.3  & Subaru HSC $^{2}$ \\
$i$ & 7711  & 1574  & 27.0  & Subaru HSC $^{2}$ \\
$z$ & 8898 & 766  & 26.4  & Subaru HSC $^{2}$ \\
$y$ & 9762 & 783  & 25.6  & Subaru HSC $^{2}$ \\ \hline \hline
\end{tabular}

Notes:\\$^{1}$ Limiting magnitude estimated with $2^{\prime\prime}$ diameter aperture.\\
$^{2}$ Subaru HSC limiting magnitudes in PDR1 are for point sources.\\

\end{table}

\subsection{Parent sample}\label{2.2}

\begin{figure}
  \centering
  \includegraphics[width=0.4\textwidth]{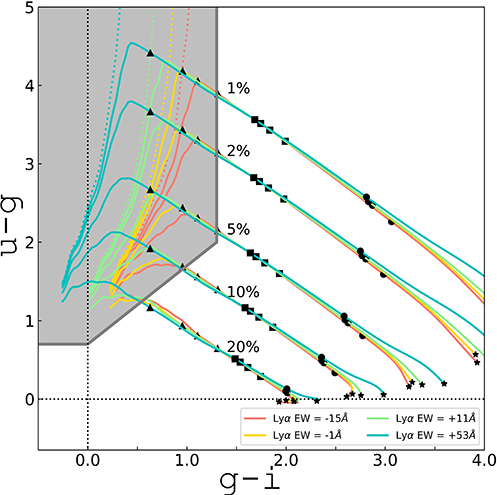}
  \caption[]{Colour-colour $u-g$ vs $g-i$ plot, the LBG selection region for $z\sim 2.7-3.4$ galaxies is shown in grey. LBG galaxies at $z>3.4$ are expected to reside off this plot using the standard LBG selection criteria. The presence of $z>3.4$ galaxies on this plot suggests the presence of LyC flux. Thus $u-g$ vs $g-i$ colur-colour plane provides a good means to identify LyC leaking galaxies. Doted blue, green, yellow and red lines are evolutionary tracks of the four composite spectra adopted from \cite{Shapley2003} with different EW(Ly$\alpha$) and without LyC flux. Evolutionary tracks after adding LyC flux to composite spectra in different ratios ($R_{\rm obs}=1\%,2\%,5\%,10\%$ and $20\%$, $R_{\rm obs}$ is defined in Section \ref{3}) are shown by solid line curves. The expected redshifts are marked in black, where triangles, squares, circles and stars correspond to $z\sim3.5, 4, 4.5$ and 5 redshift respectively. }
  \label{c-c_empty}
\end{figure}

To date, high-redshift galaxies with escaping LyC photons have usually been selected using the Lyman-break technique. 
It is recognised that during LBG selection a significant number of the galaxies are missed \citep{STEIDEL1999, LeFEVRE2005, COOKE2014}. 
Combinations of the U, G, R, and I filters are typically used to select galaxies in the redshift range $\sim 2.7<z<3.4$.
Galaxies with redshifts above $z\geq3.2$ will have $U-G$ $>$ 4, and beyond $z\sim3.4$ without any flux in the $U$ filter will, by definition, have an infinite $U-G$ colour (Figure \ref{c-c_empty} dotted lines).                  
\cite{COOKE2014} find that by examining the colours of the galaxies at $z\geq3.4$ on the $U-G$ vs $G-R$ plane we can expect to find galaxies that emit Lyman continuum flux at redshifts $z>3.5$, located outside standard LBG box (solid lines Figure \ref{c-c_empty}).
Driven by that idea, we will focus our investigations on those galaxies that are not classified as LBGs (outside the standard LBG selection region) and that are at $z\geq3.5$.
This approach allows us to test predictions from \cite{COOKE2014}.

In this work, we utilize spectroscopic redshifts from several surveys in the literature that cover the COSMOS field.
By selecting objects with redshifts $z\geq3.5$ that have high quality flags, determined by different surveys, we ensure that any flux detection in the CLAUDS $u$-band imaging is a clean LyC detection.  This means that the flux observed for the object is not contaminated by the non-ionizing radiation.

To create a parent sample that will be examined in Section \ref{2.3} for LyC radiation, we first select all objects with spectroscopic redshifts $z\geq3.5$ from the D10K, VUDS and 3D-HST catalogues.  
The selected objects have high quality spectrum flags that correspond to $\geq75\%$ probability that the reported redshift is correct.

\begin{figure}
  \centering
  \includegraphics[width=0.5\textwidth]{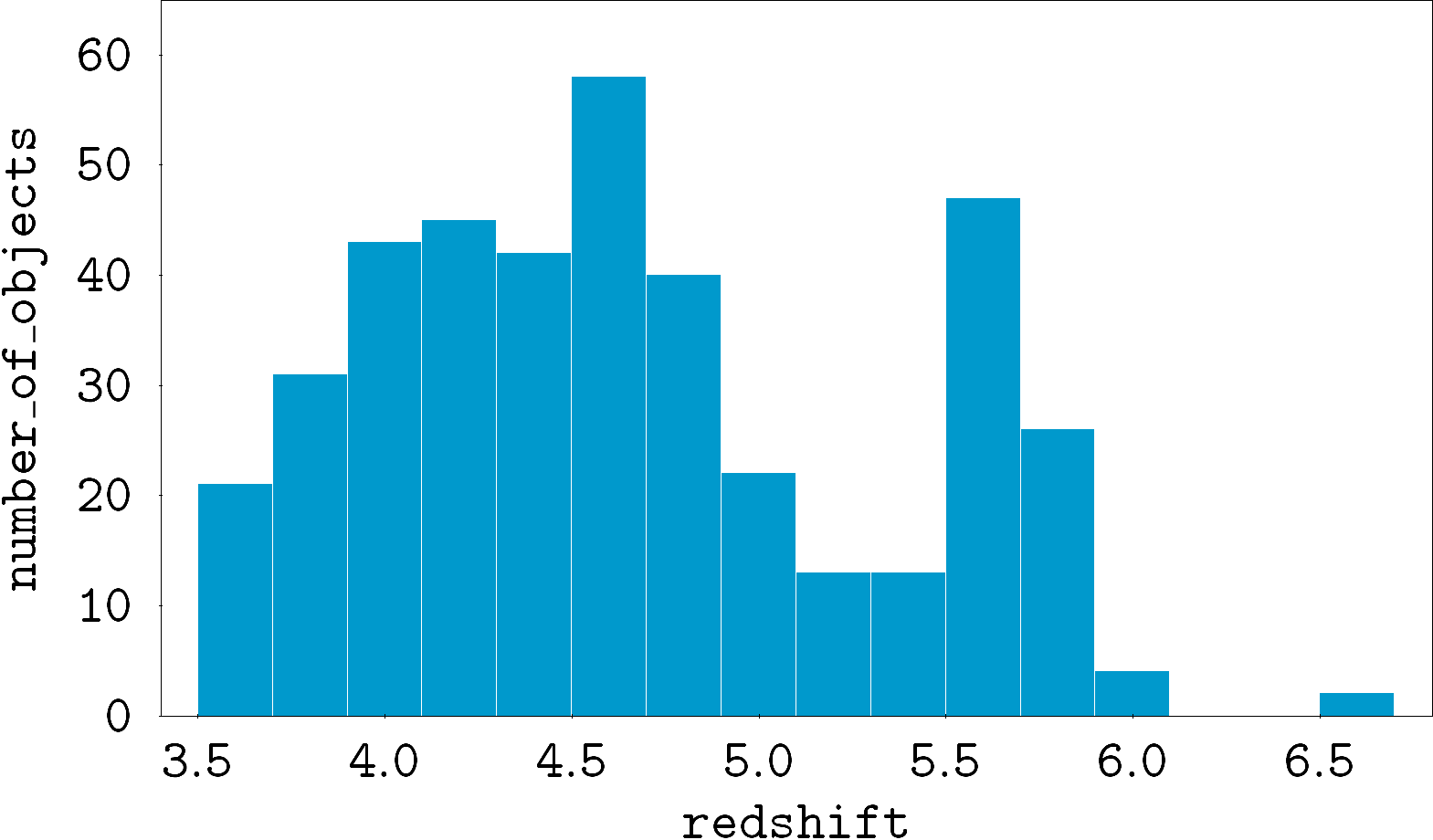}
  \caption{Redshift distribution of the parent sample, where the width of the bins is $\Delta z=0.2$. The peak at $z\sim5.7$ is due to narrow-band LAE selected sample.}
  \label{fig2}
\end{figure}

The initial selection results in 407 objects and this parent sample is presented in Table \ref{tab2} and the redshift distribution is shown in Figure \ref{fig2}.
Most of the redshifts in our parent sample, 361 objects (contributing 89$\%$ to the total sample used here) with $z\geq3.5$, are from the D10K survey.
The D10K survey selects objects from a variety of input catalogues based on multi-wavelength observations and, importantly, have different selection criteria.
The full D10K survey uses multi-slit spectroscopy that covers the wavelength range $\sim$ 5500\AA\ -- 9800\AA\ and objects are identified up to $z\sim6$.  
More details on the observations, target selection and reduction can be found in \cite{Hasinger2018A}.
From the VUDS spectroscopic survey we were able to extract 40 objects (10\% of the total sample) that have $z\geq3.5$.  
VUDS spectroscopy covers the 3650\AA\ -- 9350\AA\ wavelength range and targets objects at all redshifts to $z\sim6$.
More details on the observations, target selection and reduction can be found in \cite{LeFevre2015}.  
From the available grism data that are a product of the 3D-HST survey, we extract 6 objects (1\%) with $z\geq3.5$.
For more details about 3D-HST grism spectroscopy, we refer the reader to \cite{Momcheva2016}.

Next, the parent sample is cross-matched with the Subaru HSC catalog to obtain photometry of the sources within $0.5^{\prime\prime}$ using the \textsc{topcat} software \citep{TOPCAT}, resulting on 375 object matches (32 objects not matched).
The aperture centres for the 375 objects are defined by adopting coordinates from the Subaru HSC catalog.
For the other 32 candidates, we use the coordinates from the spectroscopic catalogs.

\begin{table}
\centering
\caption{Parent sample of galaxies with spectroscopic redshifts}
\label{tab2}
\begin{tabular}{ccc}
\hline
\hline
Survey               & Number of objects & Percentage of parent sample \\ \hline
D10k           & 361               & 89\%                \\
VUDS           & 40                & 10\%                 \\
3D-HST         & 6                 & 1\%                 \\ \hline
\textbf{TOTAL} & \textbf{407}      & \textbf{100\%}         \\ \hline \hline
\end{tabular}
\end{table}

\subsection{Sub-samples}\label{2.3}

With the parent sample now defined, the next step is to perform aperture photometry.
We perform $1.2^{\prime\prime}$ diameter circular aperture photometry with the Python \textsc{Astropy} package \textsc{photutils} designed to detect and perform photometry of astronomical sources. 
Photometry on the CLAUDS $u$-band images resulted in positive flux detection in 22 candidates with $S/N>3$ and 151 candidates with $S/N<3$.
For 234 candidates the estimated LyC flux was negative.
For this work, we will continue our analysis on the objects with reported positive flux for a total of 173 candidates (22 with $S/N>3$ and 151 with $0<S/N<3$) and from 234 candidates with negative flux we will use only candidates that show no other object inside a radius of $1^{\prime\prime}$ from the candidate.

In the next phase of the sample selection, we visually inspect all 173 candidates with positive flux. 
Thumbnails of size $15^{\prime\prime}\times15^{\prime\prime}$ are created in all available filters and all candidates are checked for LyC signal using CLAUDS $u$-band image with $\sim 0.2^{\prime\prime}/\rm pix$ resolution and median seeing of $0.80^{\prime\prime}-0.85^{\prime\prime}$ \citep{Sawicki2019}.  
The Subaru HSC images in the $g$, $r$, $i$ and $z$ filters are also inspected for flux contamination from nearby bright objects and star spikes. 
Previous research by \cite{Vanzella2010} and \cite{Siana2015} indicates that the probability of contamination by foreground objects should not be neglected and depends on seeing conditions, aperture size, limiting magnitude and increasing redshift.  
Based on the \cite{Vanzella2010} results, we would expect that the fraction of contaminated candidates in our sample would be $\sim 3 - 15 \%$ in the case of the $0.5^{\prime\prime}$ to $1^{\prime\prime}$ seeing, respectively.  
To decrease the fraction of candidates contaminated by foreground objects we use high resolution $HST$ imaging in the F125W and F160W \citep{Kokemoer2011,Grogin2011,Skeleton2014}, F140W \citep{Skeleton2014}, F336W and F435W (Prichard et all. in prep), F606W \citep{Kokemoer2011, Grogin2011} and F814W bands. 
The $HST$ F814W and F606W filter images have $\sim0.03^{\prime\prime}{\rm pix}$ resolution while images in other $HST$ filters have $\sim0.06^{\prime\prime}{\rm pix}$ resolution.  
Imaging in all mentioned $HST$ filters is not available for every object due to different areas of $HST$ imaging coverage, however, F814W is available for 95.6\% (389) of our sample of 407 objects.
The same strategy of the visual inspection is adopted for the 234 candidates with negative flux to select only candidates without any source of radiation within $1^{\prime\prime}$ radius. 
We classify 53 candidates with negative flux as not contaminated by flux from the nearby galaxies or other sources of radiation, with 181 candidates are not suitable for further analysis due to external flux contamination.

For the purpose of the visual classification, the five sub-samples are defined as follows:
\begin{itemize}

  \item $\textit{Detection}$: These are the objects that are considered clean, non-contaminated LyC emitters. 
  They are defined as galaxies with no evidence of another object within a $1^{\prime\prime}$ radius and the signal of the candidate in any band ($u$, $g$, $r$, $i$, $z$ or any available $HST$ band) is not contaminated by flux leakage from nearby bright objects or star spikes and there is no evidence of an intervening object in the spectra. We identify 2 objects ($0.5\%$ of our parent sample) in this sub-sample with $S/N\geq3$.

  \item $\textit{Detection close pairs}$: These are objects where the candidate shows LyC flux but another object is located within a $1^{\prime\prime}$ radius and does not appear to contaminate the LyC radiation from candidate. We identify 5 objects ($1.2\%$ of our parent sample) in this sub-sample with $S/N\geq3$.

  \item $\textit{Non-detection}$: These are non-contaminated objects (i.e., no object is detected within a $1^{\prime\prime}$ radius) for which LyC radiation is not detected.  We identify 35 objects with positive measured flux and 53 objects with negative measured flux, in total 87 objects ($21.6\%$ of our parent sample) in this sub-sample.

  \item $\textit{Multiple objects}$: These are the objects that appear to be more than one galaxy within a $1^{\prime\prime}$ radius and without LyC flux from the candidate, and are either mergers or are contaminated by low redshift interloper(s). 
  It is difficult to distinguish the two types from images alone.
  By default we include any object without HST coverage in this group, as we are not able to determine whether the source is contaminated by another object within the $1^{\prime\prime}$ radius.  Objects from this sub-sample will not be considered further in this work.  We identify 118 objects ($29\%$ of our parent sample) in this sub-sample.

  \item $\textit{Flux contaminated}$: These are the objects that are contaminated by flux from nearby bright galaxies, stars or star spikes in any of the examined filters.  Objects in this sub-sample will not be used for any kind of analysis.  We identify 13 objects ($3.2\%$ of our parent sample) in this sub-sample.

  \item $\textit{Negative flux - contaminated}$: These are the objects that appear to be more than one galaxy within a $1^{\prime\prime}$ radius and measured flux from the candidate is negative. We identify 182 of these objects ($44.7\%$ of our parent sample)
  
\end{itemize}

The final outcome from visual classification of the parent sample is summarized in Table \ref{tab3} where all candidates in the \textit{Detection} and \textit{Detection close pairs} sub-sample have S/N>3 and all non detection have S/N<3.

\begin{table}
\centering
\caption{Outcome from visual classification of the parent sample. This list is subject to further spectral confirmation check as described in section \ref{4}.}
\label{tab3}
\begin{tabular}{ccc}
\hline
\hline
Sub-sample  & Number of objects & percentage \\ \hline
detection             & 2  & 0.5\%\\
detection close pairs & 5  & 1.2\% \\
non-detection         & 87  & 21.4\% \\
multiple objects      & 118 & 29\% \\
flux contaminated    & 13  & 3.2\% \\ 
negative flux - contaminated & 182 & 44.7 \% \\ \hline
\textbf{TOTAL} & \textbf{407}      & \textbf{100\%}         \\ \hline \hline
\end{tabular}
\end{table}

\section{Spectral confirmation}\label{4}

With the methodology now described we will proceed with a description of carefully inspecting the sub-samples, with the goal of creating a non-contaminated sample that is as clean as possible from low redshift objects.  
We inspect the 1D and 2D spectra (if available) to double check the reported spectroscopic redshifts and carefully examine the spectrum for any possible foreground contamination.  
The slit position (if available) of the candidate is also compared with the position of the dispersion line in the 2D spectrum to confirm that the reported redshift belongs to the candidate object.  
Furthermore, to help rule out the possibility that our candidates are AGNs/QSOs, the spectra are examined for AGN far-UV emission-line signatures and $\textit{Detection}$ and \textit{Detection close pairs} are cross-matched with the publicly available XMM Newton X-ray \citep{Cappelluti2009} and \textit{Chandra} COSMOS legacy \citep{Marchesi2016} catalogues within $0.5^{\prime\prime}$.
The results of the spectral confirmation are summarized in the Table \ref{tab4} and described in the following subsections \ref{4.1} and \ref{4.2}, where two quality groups are created.
For objects classified as quality 1 (q1) we were able to confirm their reported redshifts and for quality 2 (q2) we were unable to solidly confirm reported redshifts.

\subsection{\textit{Detection sub-sample}}\label{4.1}

After visual selection and spectroscopic examination, our $\textit{Detection}$ sub-sample contains 2 candidates that show possible non-contaminated LyC flux.
For both candidates 1D and 2D spectra are re-inspected to confirm redshifts and rule out possible contamination by low redshift interlopers.
The equivalent width for all candidates are measured in the same way as described in \cite{Cassata2015} using the \textsc{IRAF} \citep{Tody1986} \textsc{splot} tool from \textsc{noao.onedspec} package.

By cross-matching coordinates with the XMM Newton X-ray and Chandra legacy catalogue we did not characterize either candidate as an AGN.
Individual analysis of the available 1D and 2D spectra from the literature, as well as slit positions and multi-band imaging data revealed the following:
\begin{itemize}
  
  \item Candidate id 394: the 1D and 2D spectra from the VUDS survey were available for analysis. 
  Analysing 1D and 2D spectra we were unable to claim any spectroscopic feature that would indicate reported redshift, so we exclude this candidate from further analysis. 
  
  \item Candidate id 421: the spectrum from the VUDS survey was available for analysis. 
  From the spectrum we were able to identify several spectroscopic features consistent with the reported redshift from the VUDS survey, such as a Ly$\alpha$ forest break, Ly$\beta$ absorption, and potential ISM absorption features, but conservatively cannot claim a solid redshift confirmation.
  The spectrum is presented in Appendix B.

\end{itemize}

\begin{figure}
\begin{center}
    \subfloat{\includegraphics[width=.48\textwidth]{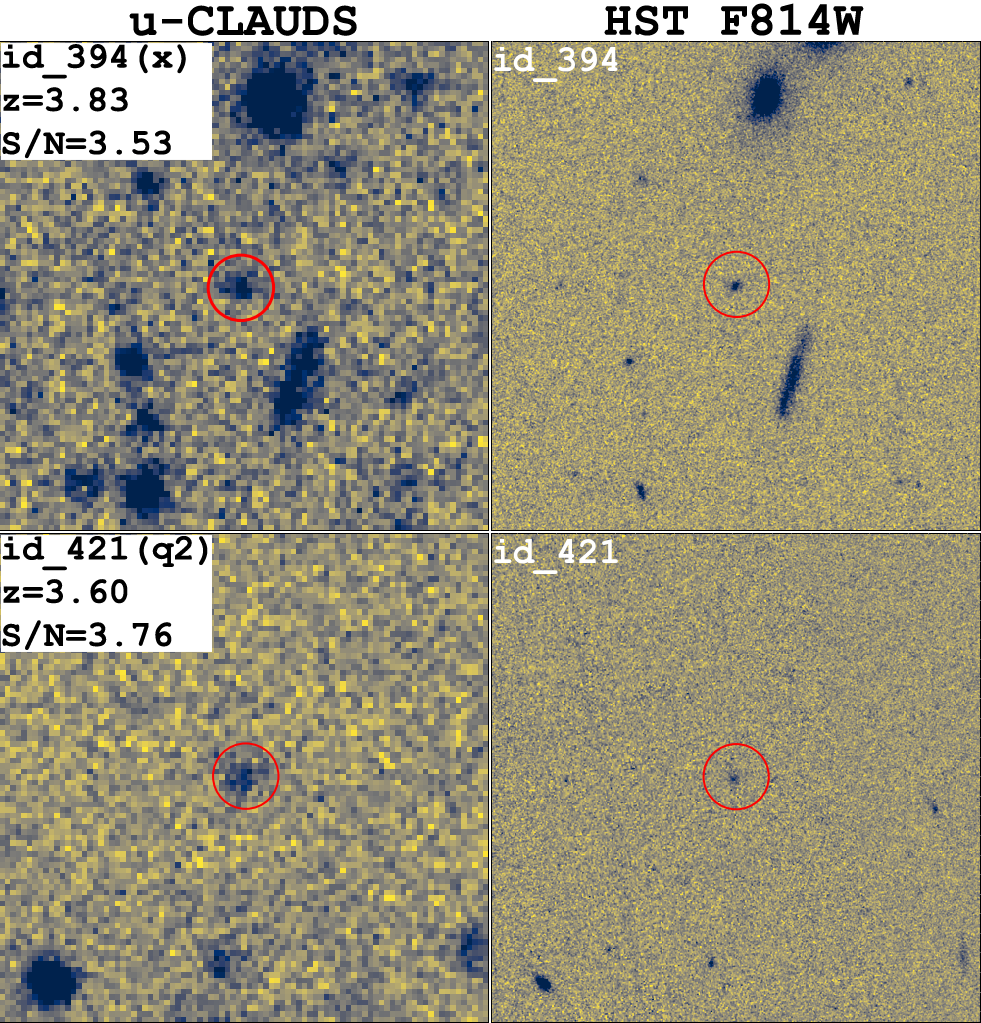}}
\caption{Thumbnails $15^{\prime\prime}\times15^{\prime\prime}$ in size for the $\textit{Detection}$ sub-sample. Candidates are shown in two bands: CLAUDS-$u$ and HST F814W. 
The red circle is $2^{\prime\prime}$ diameter aperture. Next to the id number of the candidate in brackets are quality group of the objects (q1 or q2) or x which means that candidate is rejected.}
\label{det1_2}
\end{center}
\end{figure}

In Section \ref{5} we will refer to the object id 421 as a second quality (q2) candidate.  Thumbnails of the $\textit{Detection}$ sub-sample in the CLAUDS $u$ and HST F814W bands are presented in Figure \ref{det1_2}.

\begin{table*}
\caption{Properties of the candidates after spectral confirmation procedure. Listed magnitudes in $u$, $i$ and $z$ bands are calculated in the $1.2^{\prime\prime}$ circular aperture. }
\label{tab4}
\begin{tabular}{cccccccccccl}
\hline
\hline
id  & \begin{tabular}[c]{@{}c@{}}RA\\ (deg)\end{tabular} & \begin{tabular}[c]{@{}c@{}}Dec\\ (deg)\end{tabular} & \begin{tabular}[c]{@{}c@{}}$z_{\rm spec}$\end{tabular} & \begin{tabular}[c]{@{}c@{}}$u$\\ (mag)\end{tabular} & \begin{tabular}[c]{@{}c@{}}$i$\\ (mag)\end{tabular} & \begin{tabular}[c]{@{}c@{}}$z$\\ (mag)\end{tabular} &\begin{tabular}[c]{@{}c@{}}S/N\\($u$-band)\end{tabular} &\begin{tabular}[c]{@{}c@{}}EW(Ly$\alpha)_{rest}$\\ (\AA) \end{tabular} &\begin{tabular}[c]{@{}c@{}}quality\\ (q) \end{tabular} &\begin{tabular}[c]{@{}c@{}}sub-sample \end{tabular} \\  \hline
1   & 149.596  & 2.269  & 4.28   & $27.80\pm0.33$  & $25.42\pm0.03$  & $25.17\pm0.04$  & 3.07  & $48\pm10$  & 1 & close pairs \\
326 & 150.403  & 1.879  & 3.57   & $27.10\pm0.23$  & $25.12\pm0.03$  & $24.77\pm0.04$  & 4.33  &  $\leq0$   & 2 & close pairs \\
330 & 150.443  & 1.992  & 5.09   & $27.57\pm0.29$  & $25.71\pm0.04$  & $25.55\pm0.07$  & 3.46  &  $60\pm20$ & 2 & close pairs \\
368 & 150.062  & 2.423  & 3.64   & $27.82\pm0.33$  & $25.49\pm0.04$  & $25.26\pm0.05$  & 3.05  &  $25\pm5$  & 1 & close pairs \\
421 & 150.155  & 2.413  & 3.60   & $27.40\pm0.27$  & $25.50\pm0.03$  & $25.45\pm0.06$  & 3.76  &  $\leq0$   & 2 & detection    \\ \hline \hline
\end{tabular}
\end{table*}

\subsection{\textit{Detection close pairs sub-sample}}\label{4.2}

For the $\textit{Detection close pairs}$ sub-sample the same analysis procedure is applied as for the $\textit{Detection}$ sub-sample. 
Thumbnails of the $\textit{Detection close pairs}$ sub-sample in the CLAUDS-$u$ and HST F814W bands are presented in Figure \ref{cp1_2_3}.
We visually classify 5 objects as $\textit{Detection close pairs}$ that potentially show LyC flux.  
Individual analysis of the available 1D and 2D spectra from the literature, as well as slit positions and multi-band imaging data, revealed the following information on the \textit{Detection close pairs} candidates:
\begin{itemize}

  \item Candidate id 1 has features of a Ly$\alpha$ emitter, namely a relatively strong, asymmetric emission feature in the 1D and 2D spectrum typical of Ly$\alpha$ lines at high redshift and a higher flux level in the UV continuum redward of Ly$\alpha$ as compared to blueward (i.e., the Ly$\alpha$ forest).
  This object will be noted as first quality (q1) candidate.

  \item Candidates id 326 and id 330 have somewhat asymmetric emission lines, consistent with those observed for Lyman break galaxies, and are probably Ly$\alpha$ (see Appendix B). 
  The continua are faint or not detected in the spectra.
  For these candidates we were not able to securely confirm their reported redshifts and in Section \ref{5} these objects will be noted as second quality (q2) candidates.
  
  \item Candidate id 368 has detected Ly$\alpha$ with rest frame EW(Ly$\alpha)\sim25$\AA.  The Ly$\alpha$ spectral profile, appropriately strong Ly$\alpha$ forest break, UV continuum profile, and ISM line absorption features are consistent with a  $z=3.64$ star forming galaxy.  
  Further inspection of the spectra did not reveal any other lines that could be related to low redshift objects.
  Candidate id 368 was part of the clean sample in \cite{Marchi2017} where they estimate the LyC signal from galaxies at $z\sim4$ using VIMOS spectroscopy and available HST imaging. 
  Furthermore, \cite{Marchi2017} do not report any single solid detection of individual candidates. 
  On the contrary, in this work using CLAUDS $u$-band photometry we detect a LyC signal from candidate id 368, and its $z=3.64$ has now been additionally confirmed with MOSFIRE spectroscopy where the nebular emission lines [OIII] doublet and H$\beta$ are detected (Bassett et al., in prep). 
  Beside nebular lines, the MOSFIRE spectrum reveals no other lines, which additionally confirms that detected LyC signal is not contaminated by low redshift interlopers.
  VUDS and MOSFIRE spectra are shown in the Appendix B.
  In Section \ref{5}, we will refer to the candidate id 368 as a first quality candidate (q1).

  \item Candidate id 359 has two or more independent emission lines indicating a merger system or contamination by a low redshift object and both objects fall in the slit.
  We label the bottom object "object A" and upper object "object B" in Figure \ref{cp1_2_3}. 
  In the 2D spectrum the two emission lines are visible, as well as two faint continua which is in agreement with the positions of the objects in the slit. 
  From the 2D spectrum, we conclude that the line at $\sim7747$\AA\ is related to object A, whereas the line at $\sim8167$\AA\ is related to object B. 
  If both detected lines are Ly$\alpha$ the estimated redshifts would be $z\sim5.37$ and $z\sim5.72$ for objects A and B respectively. 
  Since the spectrum has a low S/N and the two objects are to close to each other it is difficult to extract them separately and to check for the expected drop in flux blue-ward of the Ly$\alpha$ line. 
  As a result, we choose to to remove this candidate from the further analysis.
  
\end{itemize}

\begin{figure*}
  \centering
  \includegraphics[width=10cm, height=11cm]{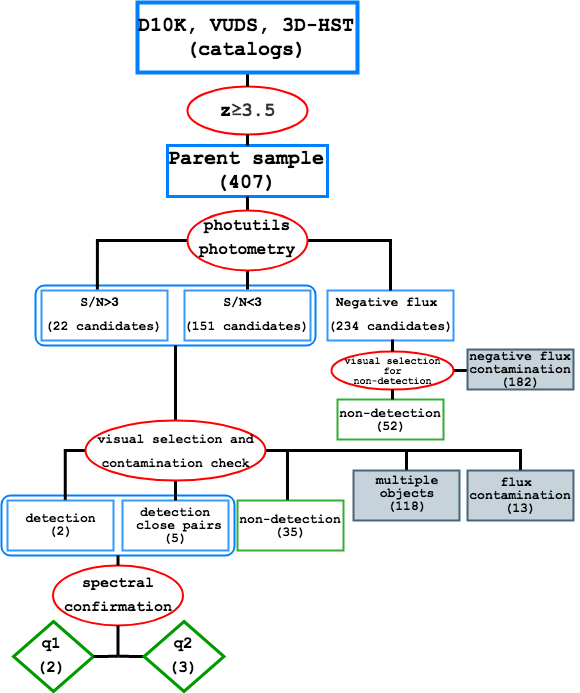}
  \caption{Flow chart summarizing the steps of the LCG candidate selection process.  The action described in each red circle applies to the sample of objects in the blue rectangle above it. Gray shaded rectangles are sub-samples that are not discussed in this work and the green diamonds and rectangle are sub-samples that are the main focus of this work (q1, q2 and \textit{Non-detection}).}
  \label{flow_chart}
\end{figure*}

\begin{figure}
\begin{center}
    \subfloat{\includegraphics[width=.48\textwidth]{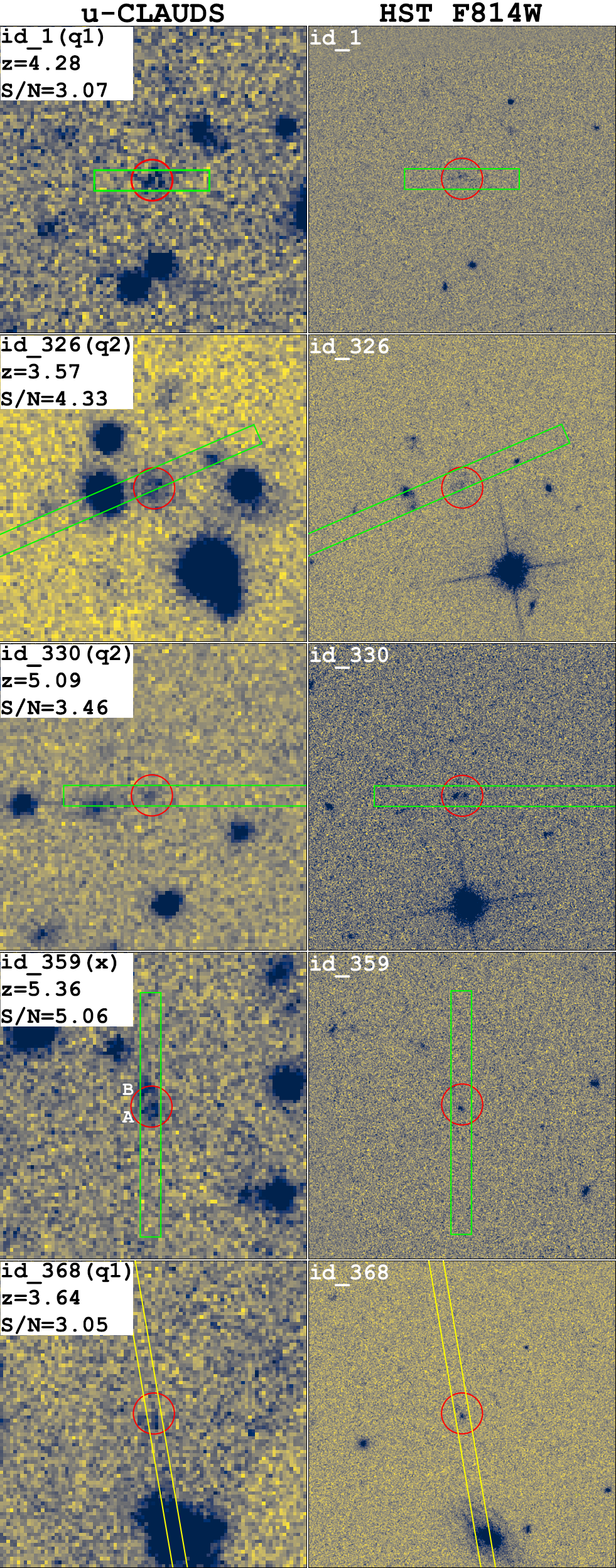}}
\caption{Same as Figure \ref{det1_2} but thumbnails are for the \textit{Detection close pairs} sub-sample. The green lines are slit positions from D10k and for object id 368, the yellow lines represent MOSFIRE slit positions with $0.7^{\prime\prime}$ width. The object with (x) next to the id number has been removed see Section \ref{4.2}.}
\label{cp1_2_3}
\end{center}
\end{figure}

To conclude this section on spectral confirmation, we provide a flow chart in Figure \ref{flow_chart} where the complete selection process of our LyC candidates is summarized.  In total, we find two q1 and three q2 candidates.

\section{Calculation and expression of the LyC escape fraction}\label{3}

One of the least constrained parameters in both observational and theoretical studies of reionization is the escape fraction of LyC photons ($f_{\rm esc}$).  
The escaping LyC photons produced by O and B stars in star-forming galaxies are referred to as the absolute $f_{\rm esc}$, or $f_{\rm esc}^{\rm abs}$, and relative $f_{\rm esc}$, or $f_{\rm esc}^{\rm rel}$.
The quantity $f_{\rm esc}^{\rm abs}$ is defined as the fraction of the ionizing photons that escape without being absorbed by interstellar medium (ISM) or circumgalactic medium (CGM) into the IGM.  
Direct measurements of $f_{\rm esc}^{\rm abs}$ are not possible since the intrinsic ratio of the ionizing and non-ionizing UV photons is difficult to determine. 
Moreover, the measurement of the ionizing and non-ionizing UV radiation is severely affected and suppressed by the high IGM opacity towards higher redshifts and dust attenuation.  
We therefore have to model these missing parameters.  
Modeling SEDs of distant star forming galaxies by using different stellar population synthesis models and taking into account dust extinction can produce insights in the intrinsic properties of the galaxies at high redshifts. 
The quantity estimated to help observations of LyC emitters is $f_{\rm esc}^{\rm rel}$ \citep[e.g.][]{STEIDEL2001}, and is defined as:

\begin{equation}
\label{eq:1}
f_{\rm esc}^{\rm rel}=\frac{(F_{LyC}/F_{1500})_{\rm obs}}{(L_{LyC}/L_{1500})_{\rm int}}\exp{}(\tau_{IGM}^{LyC}),
\end{equation}

\noindent where $(F_{LyC}/F_{1500})_{\rm obs}$ is the observed restframe LyC to UV flux density, $(L_{LyC}/L_{1500})_{int}$ is the intrinsic ratio of the galactic ionizing (LyC) to non-ionizing (UV) luminosity density, and  $\tau_{IGM}^{LyC}$ is the redshift-dependent attenuation of LyC photons due to intergalactic neutral hydrogen along the line of sight.  
From models we assume $(L_{LyC}/L_{1500})_{int}$ varies in the range of $\sim 0.1-0.9$ and depends on several galactic parameters like star formation history, stellar initial mass function, stellar age, and metallicity.  
The attenuation factor $\tau_{IGM}^{LyC}$ is usually determined by analytic models or MC simulations \citep[e.g.,][]{Inoue2014, Steidel2018}.  
Finally, we directly measure $(F_{LyC}/F_{1500})_{\rm obs}$ in this work.

We use CLAUDS $u$-band to measure $(F_{LyC})_{\rm obs}$, corresponding exclusively to the LyC flux over the rest frame wavelength range.
To probe $(F_{1500})_{\rm obs}$, rest frame non-ionizing UV flux ($\sim$1500\AA), we use Subaru HSC photometry PDR1 in $i$, $z$ and $y$ bands, depending on the redshift of the galaxy.
For the objects in the range $3.5<z<4.5$ we use $i$ band, $z$ band is used at the range $4.5<z<5.5$, and $y$ band in the $5.5<z<6.5$ range.
The non-ionizing UV flux is measured in the same aperture size ($1.2^{\prime\prime}$ diameter) and centered on the same coordinate as for LyC flux from CLAUDS $u$-band images.
In this way, we ensure that the same parts of the galaxy are probed.
The results from equation \ref{eq:1} can be directly converted to $f_{\rm esc}^{\rm abs}$ by following the equation proposed by \cite{Inoue2005} and \cite{Siana2007}:

\begin{equation}
\label{eq:2}
f_{\rm esc}^{\rm abs}=f_{\rm esc}^{\rm rel}\times 10^{-0.4(k_{1500}E(B-V))},
\end{equation}

\noindent where $k_\lambda$ is the reddening law that describes how the chosen dust model affects the UV radiation at particular wavelengths \citep[here we will use $k_\lambda=10.33$ for a Calzetti reddening law;][]{Calzetti1997} and $E(B-V)$ is the total dust attenuation or reddening.
For the purpose of our work we use $E(B-V)$ values from the publicly available catalogue of \cite{Laigle2016}.
By correcting $f_{\rm esc}^{\rm rel}$ values for internal dust attenuation we are able to obtain rough estimates for $f_{\rm esc}^{\rm abs}$.  

The only measured value in equation~\ref{eq:1} is $(F_{LyC}/F_{1500})_{\rm obs}$.  
Another way to express this quantity is the so called relative observed fraction or $R_{\rm obs}(\lambda)$ proposed by \cite{COOKE2014}:

\begin{equation}
\label{eq:3}
R_{\rm obs}(\lambda) \equiv \frac{F_{\rm obs}^{LyC}}{F_{\rm obs}^{UV}},
\end{equation}

\noindent where $F_{\rm obs}^{LyC}$ is the observed ionizing radiation flux integrated over the filter probing the LyC (here, the CLAUDS $u$-band) and $F_{\rm obs}^{UV}$ is the observed non-ionizing UV radiation near 1500\AA\ (here, the Subaru, $i$ or $z$ bands).  
In the literature $R_{\rm obs}(\lambda)$ is usually referred to as the flux density ratio, $(f_{1500}/f_{900})_{\rm obs}$.   
The advantages of using $R_{\rm obs}(\lambda)$ in applications like this is that it is not model dependent, derived directly from the observations.  
In addition, it results in an arguably more practical value (i.e., \% LyC flux) and the typically smaller errors on the UV continuum are in the denominator.
The value $R_{\rm obs}(\lambda)$ is advantageous here, as we are using only data from filter observations and in comparison with other galaxy quantities that are also derived from the observations [e.g. magnitude, colour and EW(Ly$\alpha$)].

\subsection{Attenuation of the IGM - $\tau_{IGM}^{LyC}$}

For the purpose of this work, the mean $\tau_{IGM}^{LyC}$ is estimated by using the results from the updated analytic model for attenuation presented by \cite{Inoue2014}.
Here $\tau_{IGM}^{LyC}$ is derived as the weighted average across the whole $u$-band filter.
In that way, we are taking into the account the transmission variation of the filter.
We find that adopting the mean $\tau_{IGM}^{LyC}$ is not always the best strategy for individual objects.
In cases where we have a confirmed detection of LyC flux, the probability that this line of sight has a higher IGM transparency than the mean is not negligible. 
In these situations, using the mean $\tau_{IGM}^{LyC}$ can result in an overestimation of $f_{\rm esc}$. 
For example, the probability of a clean line of sight, where optical depth is less than unity and free from Lyman limit systems (LLSs), at 900\AA\ (source rest frame) is estimated to be $\sim70\%$ for objects at $z=3$, and $\sim20\%$ for $z=4$ \citep{INOUE2008}.

To account for the fact that any detection of LyC radiation at $z>3.6$ is likely to arise from a line of sight that is more transparent than the average transmission, we also adopt the maximum transmission ($\langle1-D_b\rangle$) estimated for the redshift range $2.7\leq z \leq5$, from \cite{Steidel2018}.
The quantity $D_{b}$ is defined as a mean depression in the rest frame continuum interval 920\AA\ - 1015\AA\ caused by Lyman line blanketing \citep{Oke1982}.
As detailed in their work, $\langle1-D_b\rangle$ is a close approximation for the maximum IGM+CGM transparency expected at given redshift that is estimated in rest frame wavelength interval 880\AA $\leq \lambda_{rest} \leq$ 900\AA.
The caveat to this approach is, for our objects, that we are probing the LyC region shortward of 880\AA.
Thus we are assuming that ionizing radiation shortward of 912\AA\ is more or less equally affected by IGM+CGM.
In this case adopting $\langle1-D_b\rangle$ as a correction factor for IGM attenuation will give us lowest possible predictions for $f_{\rm esc}$.

In this work, we are dealing with single lines of sight and estimating the actual value of $\tau_{IGM}^{LyC}$ for each case is extremely complicated and beyond the scope of this paper.  This topic will be discussed in a forthcoming paper by Bassett et. al. (in prep).
Here, we adopt these two stochastic extremes for IGM transmission, the mean $\tau_{IGM}^{LyC}$ and $\langle1-D_b\rangle$, to bracket the upper and lower limits for $f_{\rm esc}$, respectively, with each limit being an unlikely case.

\subsection{The intrinsic luminosity ratio $(L_{LyC}/L_{1500})_{int}$}

The intrinsic luminosity ratio is the most poorly constrained quantity in Equation~\ref{eq:1}.
Since we are unable to put solid constraints on the intrinsic luminosity of the galaxy from direct observations, in most cases in the literature the $(L_{LyC}/L_{1500})_{int}$ ratio is estimated from using stellar population synthesis models.
In this work, we adopt results from Binary Population and Spectral Synthesis models BPASSv2.2 \citep{Stanway2018}.
To estimate $(L_{LyC}/L_{1500})_{int}$, we use the default imf135\_300 binary population single instantaneous burst model over the mass range 0.1 - 300\(M_\odot\)\footnote[1]{The adopted IMF is based on the \cite{Kroupa1993}.} with three different metalicities sub-solar, solar and super-solar (zem5, z020, z040) for more details see \cite{Stanway2018} and the BPASS manual\footnote[2]{https://drive.google.com/file/d/1ImqPuFTYLQ7k/view}.
We calculate the $(L_{LyC}/L_{1500})_{int}$ quantity for every single candidate separately, as it depends on the time since the onset of star formation.
For this purpose, we are using three different types of synthetic spectra with sub-solar (zem5), solar (z020) and z040 metallicities in the $10^6 - 10^9$ age span.
We first convert the flux of the BPASS model synthetic spectra from solar luminosity per angstrom to luminosity densities (erg s$^{-1}$Hz$^{-1}$). 
We then normalize the synthetic extreme ultraviolet spectral region (200\AA\ - 1750\AA\ ) by the mean flux value in the range 1450\AA\ - 1525\AA.
Finally, the value of $(L_{LyC}/L_{1500})_{int}$ is estimated based on the coverage of the $u$-band filter for every candidate.
For example, for a LyC candidate at $z\sim3.6$, we take the average value in the 668\AA\ - 868\AA\ range (rest frame of the candidate) corresponding to the FWHM of the $u$-band filter for that redshift.

\section{Results and discussion}\label{5}

From here on, we present our LyC candidates in two quality groups q1 (2 objects) and q2 (3 objects) for a total of 5 objects.
The quality group assignment is based on the quality of the candidates in the images, their spectroscopic redshift, and lack of low redshift contamination evidence (\S~\ref{4}). 
Table \ref{tab_det_cp} presents the observed $R_{\rm obs}$, the rest frame LyC wavelength coverage by the CLAUDS $u$-band, together with the IGM properties $\tau_{IGM}^{LyC}$ and $\langle1-D_b\rangle$ adopted from models and estimated ranges for escaping ionizing radiation $(f_{\rm esc}^{\rm abs})$. 
These 5 LyC candidates result from the careful spectral confirmation of the sample described in Section \ref{4} of the $\textit{Detection}$ and $\textit{Detection close pairs}$ sub-samples. 
To estimate the amount of LyC flux that escapes from each candidate, we apply the described methods in Section \ref{3}.  
The estimated $f_{\rm esc}^{\rm abs}$ values with adopted IGM transmission properties are presented in Figures \ref{fesc_meanIGM} and \ref{fesc_highIGM} and discussed in the following subsections.
Thumbnails for the q1 and q2 candidates in other available HST and ground based bands are presented in Appendix A and the spectra are presented in Appendix B.

\begin{figure}
  \centering
  \includegraphics[width=8cm, height=6cm]{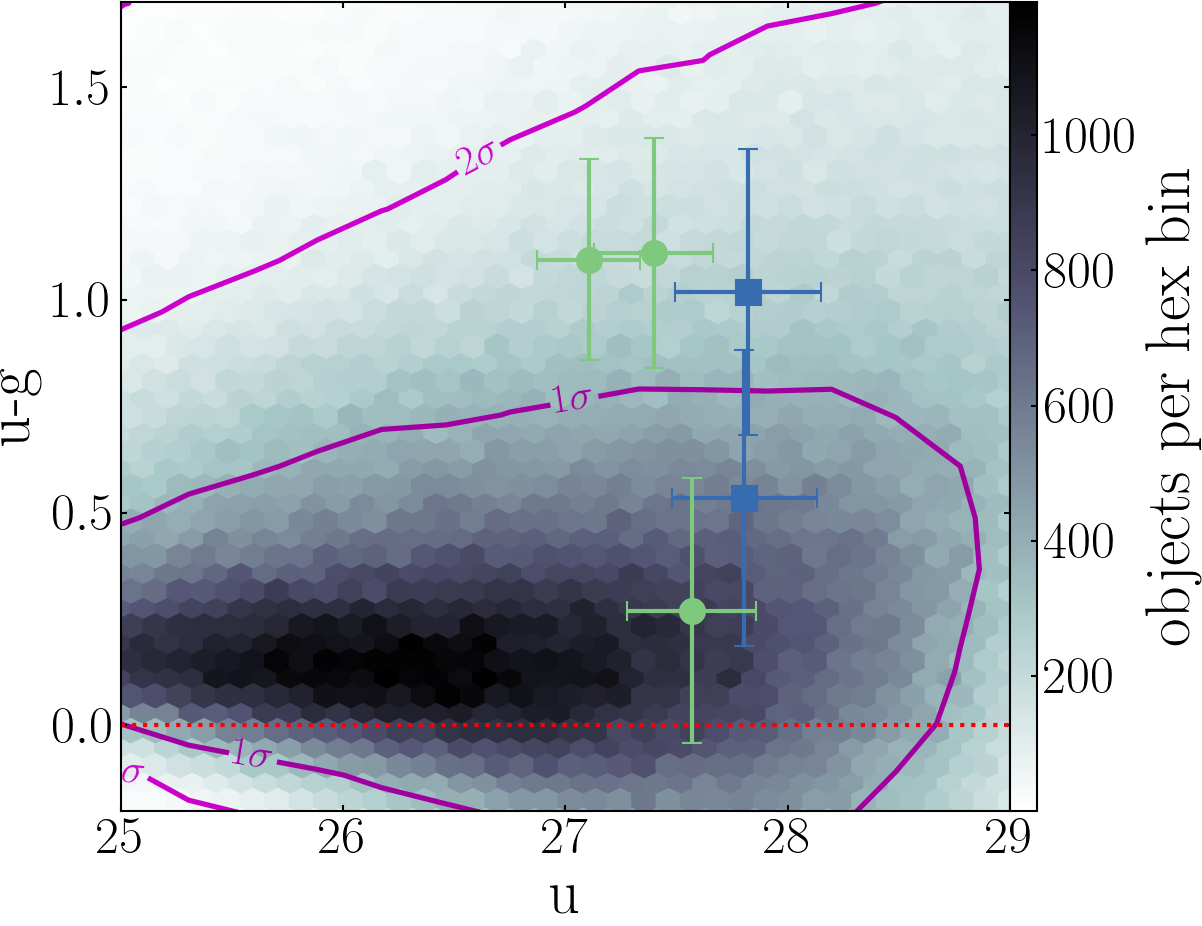}
  \caption{$u-g$ colour as a function of $u_{\rm mag}$. 
  5 LyC Candidates with q1 (blue squares) and q2 (green circles) are plotted. Colour coded hex bins present distribution of objects from the CLAUDS catalogue.}
  \label{c-mag}
\end{figure}

The candidates from both quality groups are shown in the colour magnitude diagram in Figure \ref{c-mag}, where $u-g$ colour as a function of the $u_{mag}$ is plotted.

\subsection{Estimates of $f_{\rm esc}$}\label{5.1}

From the definition of $f_{\rm esc}$, introduced in Section \ref{3}, it is clear that $f_{\rm esc}$ estimates strongly depend on the accuracy of the modeled parameters.
Under these circumstances, it is difficult to constrain the escaping flux from the candidates when there are two "free parameters" that span $0-1$ for $\tau_{IGM}^{LyC}$ and $0.1-0.9$ for $(L_{LyC}/L_{1500})_{int}$.
Because of this difficulty, we decide to estimate $f_{\rm esc}^{\rm abs}$ by using different $\tau_{IGM}^{LyC}$ and $(L_{LyC}/L_{1500})_{int}$ quantities.
As a result, $f_{\rm esc}^{\rm abs}$ values in this work are presented as a range of values, rather than as a single value.

\subsubsection{Estimates of $f_{\rm esc}$ based on the mean $\tau_{IGM}^{LyC}$}\label{5.1.1}

As discussed, $\tau_{IGM}^{LyC}$ depends on the redshift and distribution of matter along the line of sight.
The $\tau_{IGM}^{LyC}$ is estimated as a weighted average across the entire $u$-band wavelength range.
The IGM attenuation is estimated by using results from the analytical methods from \cite{Inoue2014}, where the IGM transmission is estimated by averaging over 10,000 lines of sight. 
Some of the sightlines may intersect relatively rare Lyman limit systems and, in that case, LyC photons are severely attenuated. 
An example of how LyC photons are affected when an LLS is in the line of sight and when one is absent is shown in Figure 5 of \cite{INOUE2008}.

It is important to note that the redshifts of the 5 potential LCGs (q1 and q2) reported in this work span from $z\sim3.5$ to $z\sim5$, close to the period when reionization ends.
Thus, in most cases we are probing much bluer parts of the LyC (see Table \ref{tab_det_cp}) than the 880\AA\ < $\lambda_{\rm rest}$ < 912\AA\ range most often studied in the literature.

\begin{figure*}
  \centering
  \includegraphics[width=16cm, height=11.6cm]{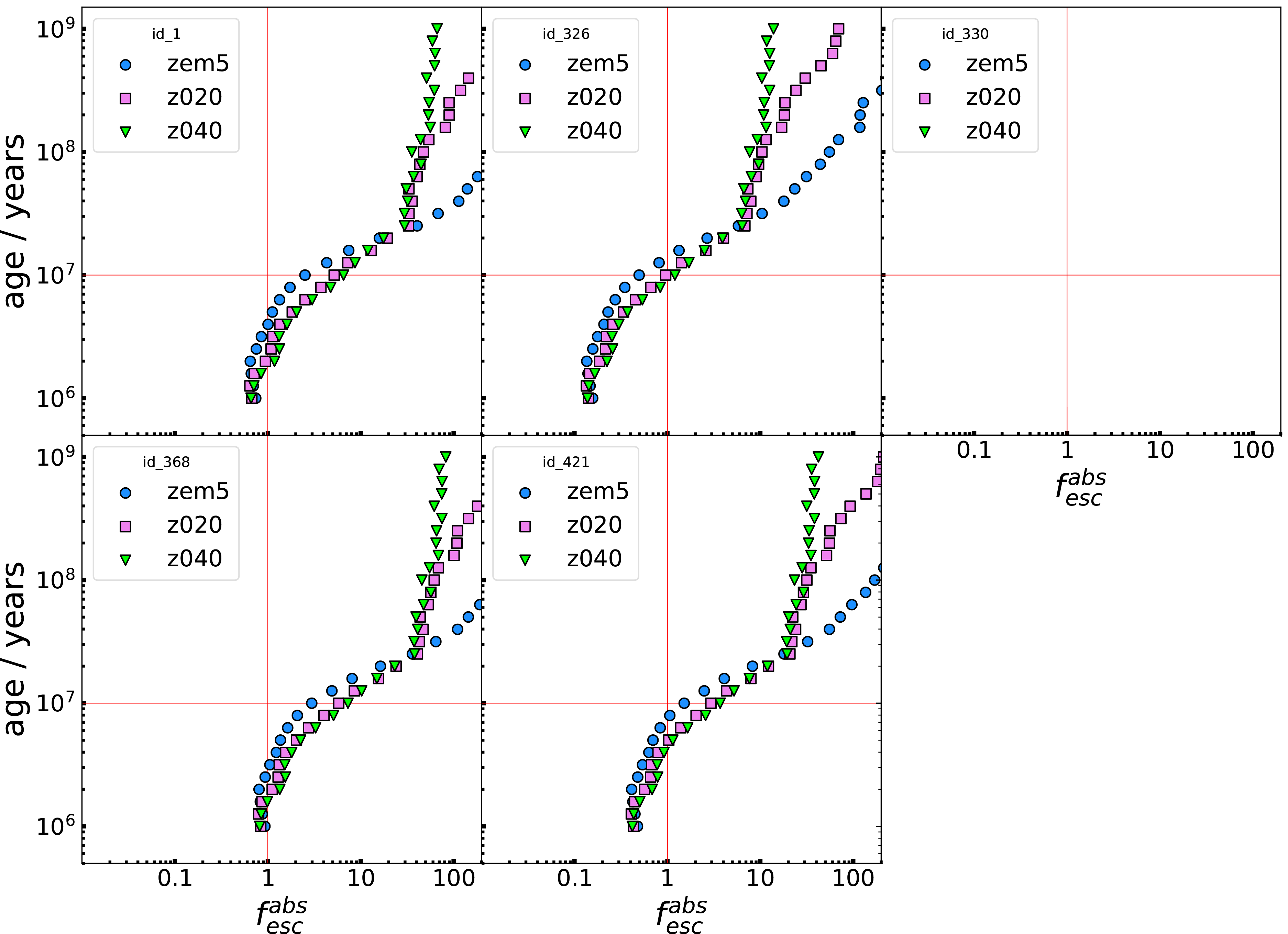}
  \caption{ The age of the stellar population plotted against estimated $f_{\rm esc}^{\rm abs}$ for our five LCG candidates, where age of the stellar population also can be interpreted as different $(L_{LyC}/L_{1500})_{int}$.  Here we adopt the mean $\tau_{IGM}^{LyC}$ from \citep{Inoue2014} (see the text for more details).  The red vertical line marks where $f_{\rm esc}^{\rm abs}=100\%$ and horizontal line shows at which point O stars end their production of LyC photons.  Blue circles, pink squares and green triangles represent the default imf135\_300 model from BPASSv2.2 with sub-solar, solar, and super-solar metallicities, respectively.  For the candidate id 330, the estimated $f_{\rm esc}^{\rm abs}$ is beyond $100\%$ (the points are outside the graph). }
  \label{fesc_meanIGM}
\end{figure*}

We estimate $f_{\rm esc}^{\rm abs}$ for our candidates after adopting the mean $\tau_{IGM}^{LyC}$ across the entire $u$-band using analytical models and $(L_{LyC}/L_{1500})_{int}$ derived from BPASS. 
The results are presented in Figure \ref{fesc_meanIGM} where the age of the stellar population, that spans from $10^6 - 10^9$ years, is plotted against estimates on $f_{\rm esc}^{\rm abs}$. 
Blue circles, pink squares and green triangles represent the same default imf135\_300 model from BPASSv2.2 with different metalicities sub-solar, solar and super-solar, respectively.
The vertical red line on the graph marks where $f_{\rm esc}^{\rm abs}$ = 100\%.
LyC radiation is mostly produced by short lived massive O-type stars, whose lifetimes are $\sim10^7$ years.
The horizontal red line marks the period when we expect the production of the LyC photons from O type stars is terminated.
As a result, the most relevant part of the plot is likely the $10^6 - 10^7$ year time range.
We estimate $f_{\rm esc}^{\rm abs}$ at different young stellar population ages, which can also be interpreted as different $(L_{LyC}/L_{1500})_{\rm int}$.
As the stellar population ages, the ratio between ionizing LyC and non-ionizing UV photons decreases. 

For the candidate id 330 $f_{\rm esc}^{\rm abs} >>100\%$.
This result can be an indication that the LyC flux from our candidate is contaminated by a low redshift interloper or that the estimated redshift is lower than $3.4$.
Another interpretation, since candidate id 330 is classified as \textit{detection-close pairs}, is that this is a merger system.
Mergers are not taken into account by BPASS population synthesis models and in these scenarios, LyC photons could be produced by fast radiative shocks, the interaction of the clumps, or accretion processes \citep{Dopita2011, Wyithe2011a}.

From Figure \ref{fesc_meanIGM}, all of the candidates show $f_{\rm esc}^{\rm abs}$ greater than $\sim0.15$ (15\%) after the O stars start to produce LyC photons.
If we assume that, at the beginning of their evolutionary paths, O stars are embedded in clouds of dust and gas, it is more likely to expect that $f_{\rm esc}^{\rm abs}$ starts from $< 0.15$ and gradually increases. 
As evolution progresses, more ionizing UV radiation is emitted and the material around the stars gets pushed away and ionized.
This can lead to the formation of the 'clean' paths or holes around systems that produce LyC photons and through these directions they can freely escape into the IGM \citep{Zackrisson2013}.

\subsubsection{Estimates of $f_{\rm esc}$ based on higher IGM transparency}

Our another approach to estimate $f_{\rm esc}^{\rm abs}$ is based on the assumption that the detected $u$-band flux in these high redshift sources indicates that our search, with its current sensitivity, is most likely biased toward lines of sight with low HI densities or free from LLSs.
Because of this, we consider whether the assumption to use the mean IGM transmission for those objects with directly detected LyC radiation is appropriate.
The influence of the IGM is also discussed by \cite{Vanzella2_2010} and their findings indicate that in some cases, transmission along the lines of sight drops to zero blueward of the redshift of the LLS, but there are cases where signal from the source is transmitted down to $\sim700$\AA\, in agreement with our observations.
If the IGM transmission of the line of sight is assumed to be higher than the mean value at a particular redshift, $f_{\rm esc}^{\rm abs}$ of the source can be smaller than the values shown in Figure \ref{fesc_meanIGM}.

\begin{table*}

\caption{Observed and modelled properties for the 5 LCGs q1 and q2 candidates.}
\label{tab_det_cp}
\begin{tabular}{ccccccccl}
\hline
\hline

id & $z_{spec}$ & $\tau_{IGM}^{LyC}$$^1$ & $\langle1-D_b\rangle$$^2$ & LyC $\lambda_{rest}$(\AA)$^3$ & R$_{\rm obs}(\lambda)$ & E(B-V)$^4$ & $f_{\rm esc}^{\rm abs}$ \\ \hline
1   & 4.28 & 0.006 & 0.334 & 568 - 763 & $0.11\pm0.03$ & 0.4  & $\gtrsim5-73\%$ \\
326 & 3.57 & 0.097 & 0.565 & 657 - 882 & $0.16\pm0.04$ & 0.3  & $\gtrsim4-15\%$ \\
330 & 5.09 & 0.00016 & 0.233 & 493 - 662 & $0.15\pm0.04$ & 0    & $>100\%$ $^5$ \\
368 & 3.64 & 0.08 & 0.565 & 647 - 869 & $0.12\pm0.04$ & 0.1  &  $\gtrsim30-93\%$\\
421 & 3.60 & 0.09  & 0.565 & 652 - 876 & $0.17\pm0.04$ & 0.2  &  $\gtrsim8-47\%$\\

\hline \hline

\end{tabular}
\\
Notes:\\$^{1}$ Mean IGM transmission estimated  from \cite{Inoue2014}.\\
$^{2}$ Close approximation to the maximum IGM+CGM transmission \citep{Steidel2018}.\\
$^{3}$ Rest frame LyC probed by CLAUDS $u$-band.\\
$^{4}$ Adopted from \cite{Laigle2016}.\\
$^{5}$ Possible low redshift galaxy or contaminated by low-z interloper. \\

\end{table*}

\begin{figure*}
  \centering
  \includegraphics[width=16cm, height=11.6cm]{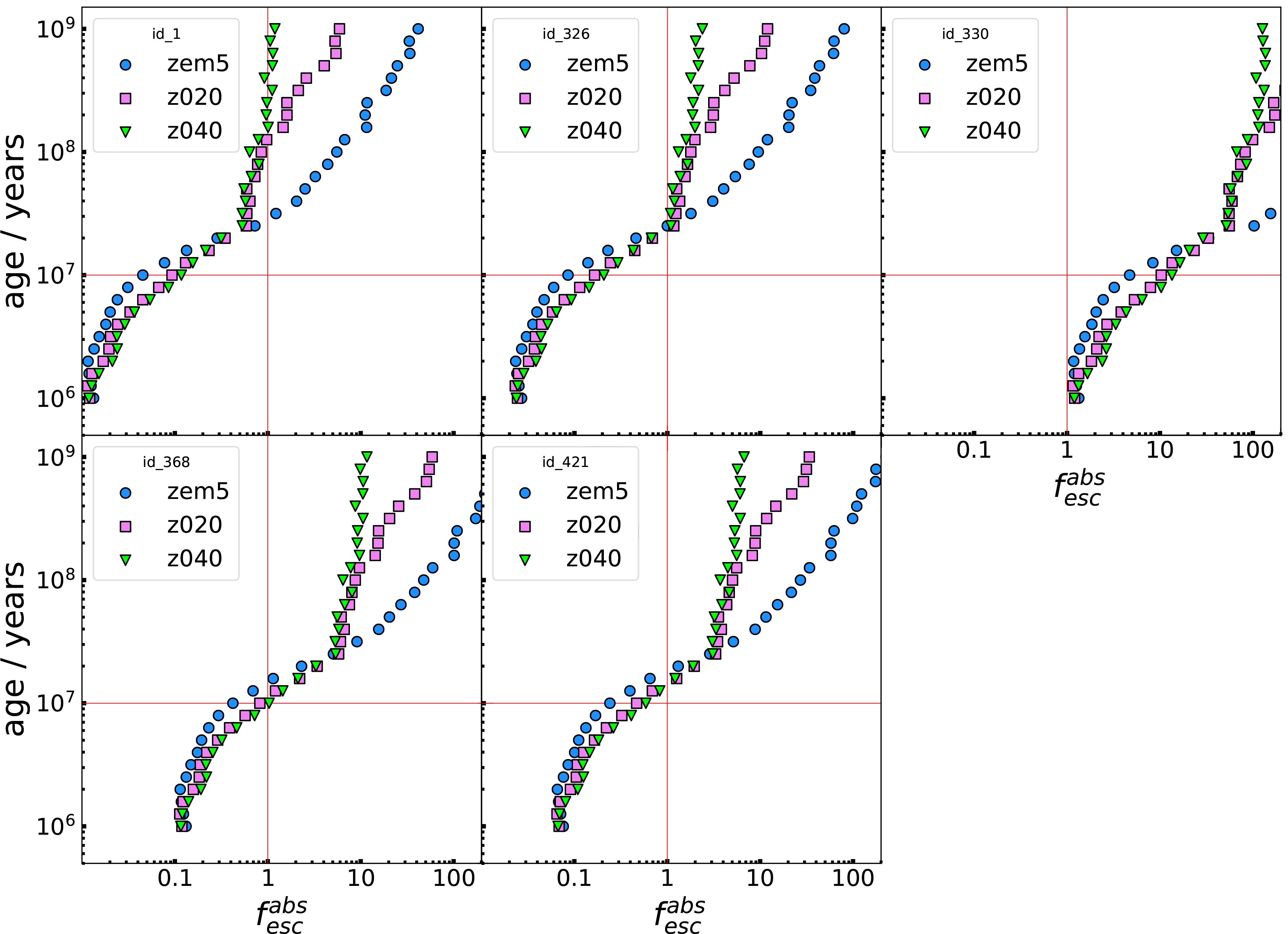}
  \caption[]{Same as Figure \ref{fesc_meanIGM}, instead of the mean $\tau_{IGM}^{LyC}$ values, we adopt a close approximation to the maximum transmission, $\langle1-D_b\rangle$ from \cite{Steidel2018} in the interval 880\AA\ $\leq \lambda_{\rm rest} \leq$900\AA.}
  
  \label{fesc_highIGM}
\end{figure*}

We have shown in Section \ref{5.1.1} that the LyC escape fraction may be overestimated if we use the mean $\tau_{IGM}^{LyC}$ to correct for IGM attenuation.
Therefore, we use IGM transmission values that describe lines of sight with higher transparency for LyC radiation.
For this purpose, we adopt the estimated maximum IGM+CGM transmission values from \cite{Steidel2018}.
A shortcoming of this approach is that the estimated IGM+CGM transmission is only available for the restframe 880\AA\ $\leq \lambda_{rest} \leq$ 900\AA\ interval and the chance for not having high column density absorber ($>10^{16}\rm cm^{2}$) at the redshifts of our candidates is less likely.
Therefore the estimated range of $f_{\rm esc}^{\rm abs}$ for our candidates can be interpreted only as lower limits.
Also for our redshift range, we only have IGM+CGM transmission estimates for discrete redshift bins $z=3.5, 4, 4.5, 5$.
The approximations of the maximum IGM transmission ($\langle1-D_b\rangle$) that we adopt for each candidate in our sample are summarized in Table \ref{tab_det_cp}.
The $f_{\rm esc}^{\rm abs}$ results based on the ($\langle1-D_b\rangle$) values are shown in Figure \ref{fesc_highIGM}, where markers and axes are the same as Figure \ref{fesc_meanIGM}.

In contrast to the previous estimates where mean $\tau_{IGM}^{LyC}$ is used, adopting the maximum transmission $\langle1-D_b\rangle$ leads to the lower $f_{\rm esc}^{\rm abs}$ values for both q1 and q2 candidates.
The candidate id 330 that belongs to the q2 sample with $f_{\rm esc}^{\rm abs}>100\%$ can be explained as a low redshift object or possible merger system as discussed in Section \ref{5.1.1}.
Therefore, by taking into account only q1 objects, we can conclude that adopting the mean IGM transmission may not properly describe the extent to which LyC photons are attenuated in individual LyC emitters (this conclusion does not change if we also include q2 objects).
On the other hand, the existence of very clean lines of sights is also less likely case at these given redshifts.
Lastly, models with different metallicities can produce variations in the estimated $f_{\rm esc}^{\rm abs}$ from a few percent in early evolutionary stages of the O stars up to $\sim10\%$ in late evolutionary stages, for both q1 and q2 candidates.

From the results presented in Figures \ref{fesc_meanIGM} and \ref{fesc_highIGM} it is clear that even with a clean detection of the LyC flux it is extremely difficult to estimate the amount of LyC photons that are leaking into the IGM, as the two parameters, $\tau_{IGM}^{LyC}$ and $(L_{LyC}/L_{1500})_{int}$ are almost impossible to measure directly and difficult to constrain with models or simulations for individual galaxies.
By not knowing at least one of the modeled parameters more precisely our calculated $f_{\rm esc}^{\rm abs}$ values are rough estimates that range from a few percent up to $\sim90\%$ (candidate id 330 excluded), in our cases, if we assume that O type stars are the main producers of LyC photons.
Therfore, as a final result we are adopting range in between upper and lower limits estimated by using $\tau_{IGM}^{LyC}$ and $\langle1-D_b\rangle$ respectively as a most plausible range for $f_{\rm esc}^{\rm abs}$ where adopted metallicity is zem5 (sub-solar), Table \ref{tab_det_cp}.

\subsection{Ly$\alpha$ properties of the LCGs}\label{5.2}

\begin{figure*}
\begin{center}
\begin{tabular}{ll}
\includegraphics[width = 0.47\textwidth]{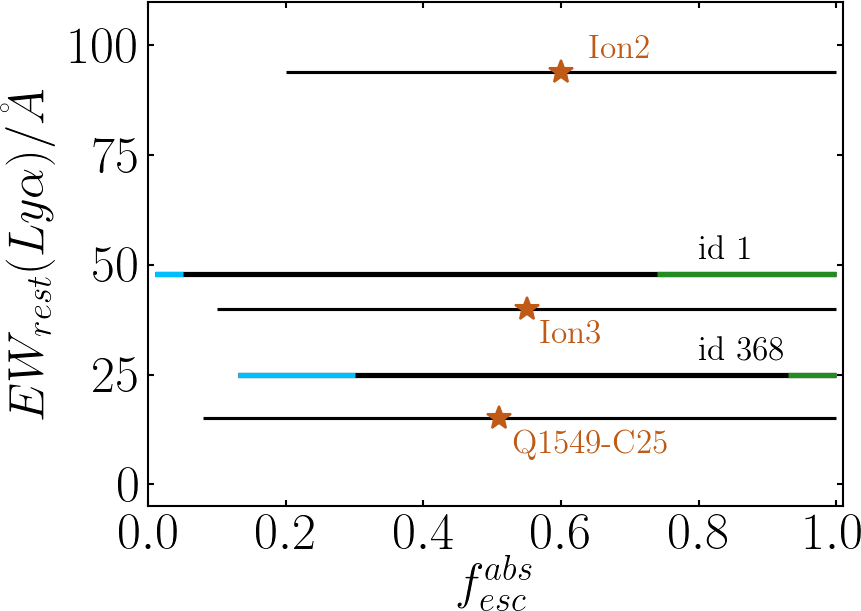} &
\includegraphics[width = 0.47\textwidth]{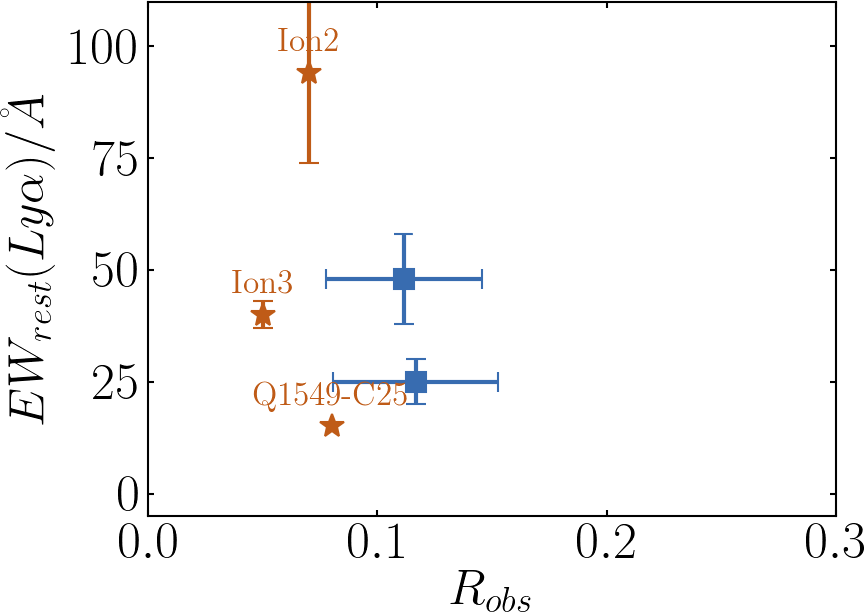} \\
\end{tabular}
\end{center}
\caption[]{$\textit{Left:}$ Rest frame EW(Ly$\alpha$) is plotted as a function of adopted $f_{\rm esc}^{\rm abs}$ range, shown as black horizontal lines.  The blue and green segments of the lines for candidates id 1 and id 368 are the $f_{\rm esc}^{\rm abs}$ ranges estimated after applying $\langle1-D_b\rangle$ and $\tau_{IGM}^{LyC}$ correction ,respectively (i.e., less likely cases).  $\textit{Left:}$ Rest frame EW(Ly$\alpha$) is plotted as a function of $R_{\rm obs}$.  The q1 candidates are marked as blue squares.  Spectroscopically confirmed LyC galaxies from literature Ion2 \citep{deBARROS2016, Vanzella2016}, Q1549-C25 \citep{Shapley2016} and Ion3 \citep{Vanzella2018} are plotted as brown stars in both left and right panels.}
\label{EWLy}
\end{figure*}

In the last few years, the link between the properties of the Ly$\alpha$ line, the escape of the Ly$\alpha$ photons, and the escape of the LyC photons among low and high-redshift galaxies has been explored by a number of independent studies.
For example, the correlation between EW(Ly$\alpha$) and $f_{\rm esc}$ among low redshift galaxies is reported by \cite{Verhamme2017, Steidel2018, Fletcher2019} and an indication of a possible trend among EW(Ly$\alpha$) and $R_{\rm obs}$ is discussed by \cite{Marchi2017}.
In Figure \ref{EWLy}, we present rest frame EW(Ly$\alpha$) as a function of estimated $f_{\rm esc}^{\rm abs}$ shown as black horizontal lines (left panel) for three spectroscopically confirmed LyC leakers and for our two q1 candidates.
Since different studies have different strategies of presenting estimated ionizing escape fractions it is important to note that for Q1549-C25 $f_{\rm esc}^{\rm abs}=51\%$ is estimated at $95\%$ confidence, where less than $45\%$ of the $f_{\rm esc}^{\rm abs}$ distribution is $\leq100\%$ and $95\%$ of the distribution is at $f_{\rm esc}^{\rm abs}>51\%$ \citep{Shapley2016}. 
In the case of the Ion2 \citep{deBARROS2016, Vanzella2016} $f_{\rm esc}^{\rm abs}$ values are described in the range from $20\%-100\%$ and for the Ion3 \citep{Vanzella2018} only $f_{\rm esc}^{\rm rel}$ is reported in range $10\%-100\%$.
Here, we are presenting $f_{\rm esc}^{\rm abs}$ for q1 candidates as a range, where we also including the estimated ranges of the less likely cases of IGM attenuation, $f_{\rm esc}^{\rm abs}$ and $\langle1-D_b\rangle$ and $\tau_{IGM}^{LyC}$, as blue and green lines, respectively.   
Although we see no clear correlation between EW(Ly$\alpha$), $f_{\rm esc}^{\rm abs}$ and $R_{\rm obs}$, with the current size of our sample we are not able to rule out its existence.
The right panel of the Figure \ref{EWLy} shows EW(Ly$\alpha$) as a function of $R_{\rm obs}$.
It is interesting to note that the q1 candidates and confirmed detections from the literature in the right panel of Figure \ref{EWLy} show an absence of strong EW(Ly$\alpha$) for higher values of $R_{\rm obs}$.

However, with current sample size and the non-uniform selection methods (D10K, VUDS, 3D-HST) in this work, as well as the fact that EW(Ly$\alpha$) line is affected by different processes (morphology, transparency of the IGM, high star formation etc.), it is not possible to rule out the existence of a correlation or anti-correlation between emitted LyC flux into IGM and properties of the Ly$\alpha$ line.
At this stage, the possible lack of the any clear trend indicates a need for a homogeneous sample selection method and larger sample.

\subsection{Position of the LCGs on the colour-colour diagram}

The Lyman break technique exploits the expected drop in flux at the Lyman limit ($<$ 912\AA) to develop selection criteria on colour-colour diagrams \citep[e.g.,][]{Steidel1996} as an efficient method of selecting $2.5\lesssim z\lesssim 3.5$ and $3.5\lesssim z\lesssim 4.5$ star forming galaxies.  These Lyman break galaxies (LBGs) have been used many previous studies searching for LyC emitting galaxies  \citep[e.g.,][]{Iwata2009, Vanzella2_2010, Nestor2011}. 
The limitations of the Lyman break selection in the context of $z\sim3-4$ LCG detection was recognised by \cite{COOKE2014}.  
The authors discuss how the effect of LyC on the u-band magnitude moves the colours of a galaxy on the colour-colour diagram from their expected LBG location for a given redshift in a predictable manner. 

Historically, the templates and composite spectra used to determine the expected positions of LBGs on colour-colour plots and, subsequently, the LBG selection regions, assumed zero LyC flux in order to use the break in flux as a selection discriminant, as a negligible fraction was expected. 
However, various levels of LyC flux enter the u-band filter and act to move the colours of the galaxies downward on a $u-g$ vs $g-i$ plot, below the locations expected for their redshifts when assuming zero LyC flux, and this effect can be sufficiently strong to place the galaxy colours outside the standard LBG selection region box.   
For LCGs with redshifts at the lower end of the redshift range being probed, the movement in colour is small (regardless of the level of LyC flux), but the movement increases (rather dramatically) toward the higher redshift end, as more LyC flux enters the $u$-band.  
In fact, for typical $u-g$ vs $g-i$ diagrams, galaxies with roughly $z\gtrsim3.3$ are positioned well off the diagram (upward in $u-g$ colour, reaching infinity for $z\gtrsim3.4$).  
Thus, the very presence of $z\gtrsim3.3$ galaxies on these plots is indicative of the presence of LyC flux.

As a result, for $z\sim3$ galaxies, inspecting their locations on conventional $u-g$ vs $g-i$ diagrams is informative regarding the potential for escaping LyC flux.  
In particular, those galaxies with colours in the region below and outside the $z\sim3$ LBG selection region (on conventional $u-g$ vs $g-i$ diagrams), as they are expected to have the highest fraction of escaping LyC flux.  
We note that this region of the diagram (below and outside the conventional selection region) contains the colours of a significant fraction of low-redshift galaxies that makes an efficient selection of $z\sim3$ LCGs difficult when using broadband optical colours alone.  

However, and perhaps more importantly, this same $u-g$ vs $g-i$ diagram provides a good means to select $z\sim4$ LCGs.  As mentioned above, LBGs at $z\gtrsim3.3$ with zero LyC flux have colours that reside off this diagram, but $z\gtrsim3.3$ galaxies with even small levels (e.g., $\gtrsim1\%$) of LyC flux are found on this plot and in locations relatively removed from the high density of low-redshift galaxy colours.
As a result, $u-g$ vs $g-i$ colour-colour plots, conventionally used for $z\sim3$ LBGs, are powerful for $z\sim4$ LCG colour selection, in particular when combined with infrared broadband colours.

\begin{figure*}
  \centering
  \includegraphics[width=17cm, height=9cm]{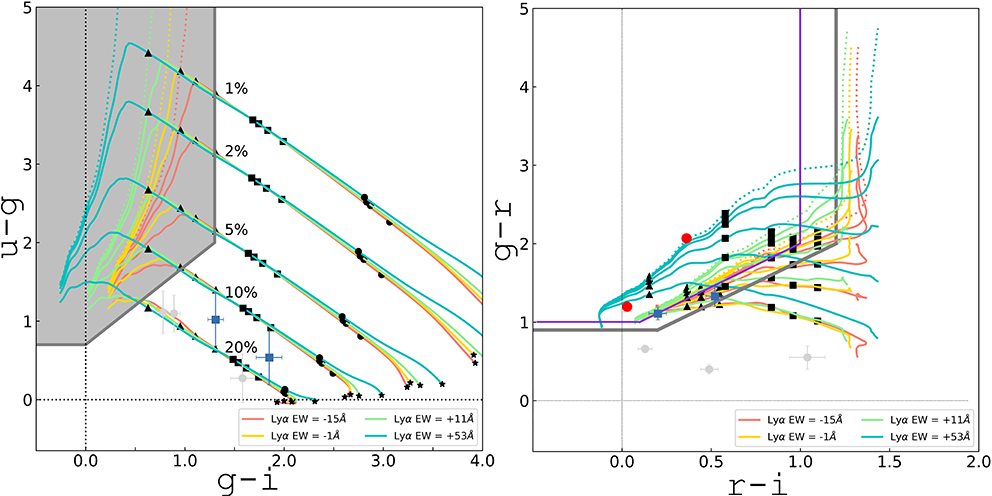}
  \caption{Colour-colour $u-g$ vs $g-i$ diagram used for selecting LCGs at redshift $\sim3-4$ (typically used for selecting LBGs at $z\sim2.7-3.4$) left and $g-r$ vs $r-i$ colour-colour diagram used for selecting LCGs at redshift $\sim4-5$ (typically used for selecting LBGs at $z>3.5$) right panel. The LBG selection region is shown in grey for $z\sim3$ LBGs. Blue, green, yellow and red colour curves are evolutionary tracks created after redshifting the four composite spectra from \citep{Shapley2003} from $z=2.7-5$.
  Dotted curves show composite spectra tracks with no LyC flux. 
  Solid line curves are also composite spectra tracks but with added LyC flux ($R_{\rm obs}=1\%,2\%,5\%,10\%$ and $20\%$) short-ward from 912\AA.
  Black triangles and squares located on the evolutionary track curves are markers for $z=3.5$ and $z=4$ respectively.
  The right colour-colour diagram has the same colour scheme as left, where black triangles and squares represent positions with redshifts $z=4$ and $z=4.5$ respectively.
  Two red circular markers represent expected positions of q1 candidates if they are free from LyC flux, while blue squares represent their positions on the diagram. The grey selection region is a more conservative one created based on evolutionary tracks and the purple region is designed based on CFHT filters (see the text for additional explanation). Blue squares on both diagrams are q1 objects and light gray circles are q2 objects.}
  \label{c-c}
\end{figure*}

The two q1 LyC candidates from this work are plotted on the $u-g$ vs $g-i$ colour-colour diagram in Figure \ref{c-c}. where the $z\sim2.7-3.4$ LBG selection region is shaded in grey.
Four dotted teal, green yellow and red curves are evolutionary tracks of four different LBG composite spectra with different EW(Ly$\alpha$) values adopted from \cite{Shapley2003}.
Here, all four composite spectra follow the standard LBG colour selection convention and assume no flux short-ward of 912\AA.
As we can see from Figure \ref{c-c}, the dotted line tracks are located inside the LBG selection region and the $u-g$ colour becomes redder as the redshift of the object increases.
The locations of the LBGs at $z\gtrsim3.3$ with zero LyC flux are above the plot and reach infinity by $z\sim3.4$. 
Adopting the flat model from \cite{COOKE2014} for all four composite spectra, which provides an average measure of the LyC flux within the $u$ filter, artificial flux is added short-ward of 912\AA\ in different amounts: $R_{\rm obs}=1\%,2\%,5\%,10\%$ and $20\%$.
Depending on their redshift and amount of ionizing radiation, we can see how the evolutionary tracks, solid teal, green, yellow and red lines, indicate that galaxies with the largest fraction of observed LyC flux shift significantly downward in $u-g$ colour, with a fraction of the $z\sim3$ galaxies residing outside the standard LBG colour selection region and the new presence of $z\gtrsim 3.3$ galaxies (black triangles and squares indicate $z=3.5$ and $z=4.0$, respectively).
If the assumption from \cite{COOKE2014} is correct then the positions of q1 candidates (blue squares) on the colour-colour diagram are consistent with this prediction and an approximate value of $R_{\rm obs}$ can be read off the plot.

We also examine positions of the two q1 candidates on the $g-r$ vs $r-i$ colour-colour diagram, right panel Figure \ref{c-c}.
The $g-r$ vs $r-i$ colour-colour diagram is designed to select LBG galaxies at redshift $3.5\lesssim z \lesssim 4.5$.
Similar to the $u-g$ vs $g-i$ colour-colour plots for selecting $z\sim3$ LBGs and $z\sim$ 3--4 LCGs, the $g-r$ vs $r-i$ colour-colour diagram is useful for selecting $z\sim4$ LBGs and $z\sim$ 4--5 LCGs.  As was the case for the $u-g$ vs $g-i$ colour-colour plots, galaxies at the lower redshift end of the redshift range probed (here, $3.5<z<4.5$) have little movement in colour, regardless of LyC flux levels, and the plot is more effective in identifying LCGs at the higher end of the redshift range and beyond to $z\sim5$.

On the right panel in Figure \ref{c-c}, the dotted evolutionary tracks of the four composite spectra are shown with no LyC flux, while solid lines are evolutionary tracks with added LyC flux.
Black triangles and squares denote the expected positions of $z=4$ and $z=4.5$ galaxy colours, respectively.
The evolutionary tracks are colour coded in the same way as in the left panel.
Here we are showing two LBG selection regions.  The purple selection region is adopted from \cite{Hildebrandt2009} and designed for CFHT $g$, $r$ and $i$ filters that have almost the same bandwidths as Subaru HSC filters.  A more inclusive selection region based on the (dotted) evolutionary tracks to $z\sim4.5$ is shown in grey. 
Our q1 candidates are marked as blue squares.  As a guide, red circles mark the positions of the colours for galaxies with similar redshifts and EW(Ly$\alpha$), but without LyC flux.
The candidates have bluer colours for their redshifts, as compared to LBGs (i.e., with zero LyC flux), and are consistent with the expectations for galaxies with measurable escaping LyC flux.  
In a similar manner to the effectiveness of the $z\sim3$ LBG $u-g$ vs $g-i$ diagram to identify $z\sim4$ LCGs, this $z\sim4$ LBG diagram is more effective in identifying LCGs at $z\sim5$.

\subsection{\textit {Non-detection} sub-sample stacking}\label{5.4}

For the candidates from the $\textit{Non-detection}$ sub-sample, we attempt to estimate the $f_{\rm esc}$ limit by stacking individual $u$-band images.
The data are stacked in two ways, based on their average and median. 
The stacking procedure is performed in \textsc{IRAF} using the \textsc{imcombine} task and flux is measured in a $1.2^{\prime\prime}$ circular aperture, the same way as was done with the single galaxy candidates.

\begin{figure}
  \centering
  \includegraphics[width=8cm]{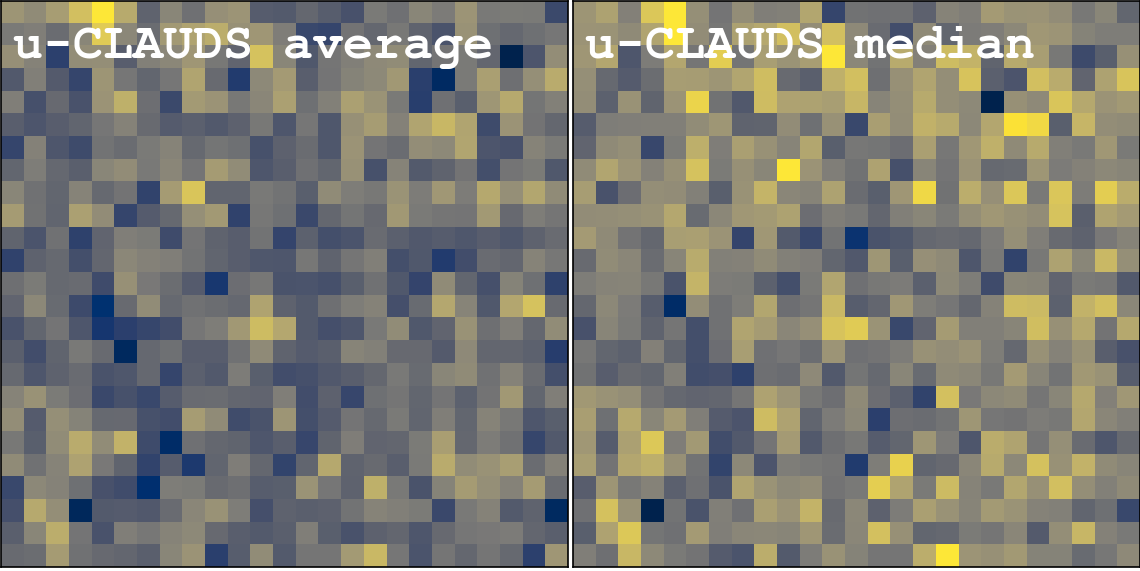}
  \caption{Stacked images of the $\textit{Non-detection}$ candidates. Thumbnails are $4^{\prime\prime}\times4^{\prime\prime}$ in size.  The average and median stacks are created by stacking 39 objects with confirmed redshifts from $\textit{Non-detection}$ sub-sample.  No signal above S/N $>3$ is detected and the estimated upper limits for $u_{avg}^{up}\sim 33.17$ mag and $u_{med}^{up}$ was negative.
  }
  \label{sn_stack}
\end{figure}

Before stacking we inspect the 1D and 2D spectra of all 87 candidates of the \textit{Non-detection} sub-sample to ensure $z>3.42$ redshifts.
Two candidates are recognized as AGN and are excluded from further analysis.
For the remaining 85 candidates, we confirm redshifts for 39.
To create stacks free from flux emitted by low redshift objects, we only use the 39 \textit{Non-detection} candidates for which we can spectroscopically confirm redshifts.
Along with the stacking of the CLAUDS $u$-band images that probe the LyC UV radiation, we stack $i$, $z$ or $y$ Subaru HSC images for non-ionizing UV flux depending on the redshift of the object.
To estimate non-ionizing UV flux for the \textit{Non-detection} sub-sample, we used the $i$-band for objects at $3.5<z<4.5$, $z$-band for objects at $4.5<z<5.5$ and $Y$-band for objects at $5.5<z<6.5$.
We create two different stacks, average and median, for each redshift bin and the non-ionizing flux is measured the same way as for $u$-band images.
This approach results in three average flux values, one for $i$, $z$ and $y$ band, and three median flux values.
Finally, to estimate the average non-ionizing UV flux we averaged the  fluxes in $i$, $z$ and $y$ band stacks.
The same is done for median values.
No LyC detection is seen with $S/N>3$ in either the average or median stack. 
The results are presented in Figure \ref{sn_stack}.

The estimated magnitude upper limit for the stack of the 39 galaxies is $u_{avg}^{up}\sim 33.17$ mag and $u_{med}^{up}$ was negative. 
In addition, we estimate the upper limits for the averaged flux density ratio $R_{\rm obs}$ and the averaged $f_{\rm esc}^{\rm abs}$ to be 0.001 and 0.006, respectively.
For the intrinsic luminosity ratio we adopt $(L_{LyC}/L_{1500})_{int}=0.3$.
The E(B-V) values for the 39 $\textit{Non-detection}$ galaxies (with confirmed redshifts) are extracted from the catalogue published by \cite{Laigle2016}.
Based on the 30 objects cross-matched by their coordinates, we adopt the average E(B-V) = 0.084 for the 39 spectroscopically confirmed $\textit{Non-detection}$ candidates.
It has been estimated that, if galaxies alone are sufficient to reionize the Universe, we would expect them to emit $\sim10-20\%$ of their LyC radiation on average into the IGM \citep{ROBERTSON2015,Khaire2016}.  The estimated upper limits on $f_{\rm esc}^{\rm abs}$ from the average and median stacks indicate that the emission of LyC radiation into IGM from galaxies is not greater than $1\%$.
From the results estimated by stacking, and their associated assumptions, we can conclude that the average amount of LyC radiation that escapes into IGM from the galaxies at $3.5<z<5.1$, if representative of galaxies at $z>6$, is not enough to sufficiently support reionization.

\section{Summary and Conclusions}\label{6}

In this work, we present the results of a search for a population of LCGs whose progenitors at $z>6$ may be responsible for reionizing the Universe.
We used deep CLAUDS $u$-band photometry in the COSMOS field to search for LyC flux emitted by LCGs. 
To ensure the cleanest sample, and that our measured flux was not contaminated by low redshift interlopers, we only include objects with high-quality spectroscopic redshifts from the D10K, VUDS and 3D-HST surveys.  HST and ground based imaging in multiple filters were examined individually for every object that potentially showed LyC flux in the $u$-band and the 1D and 2D spectra were analyzed to help eliminate low redshift contaminants next to the galaxies in the line of sight.
After the selection process, we identify 5 candidates within the redshift interval $3.5<z<5.1$, which we divide into two quality groups q1 (2 objects) and q2 (3 objects).
Our main conclusions can be summarized as follows:
\begin{itemize}
  \item{Following the approach of \cite{COOKE2014}, we use the $z\sim4$ LCG $u-g$ vs $g-i$ colour-colour diagram (conventionally used for $z\sim3$ LBG selection) to investigate our candidates.  
  The positions of the q1 LCG candidates on the $u-g$ vs $g-i$ colour-colour diagram are consistent with galaxies having measurable escaping LyC flux (here, $R_{\rm obs}\sim$ 15\%) for galaxies with their redshifts and EW(Ly$\alpha$)}.  We note that $R_{\rm obs}$ is not $f_{\rm esc}^{\rm abs}$. 
  
  \item{Adopting the mean IGM transmission values as representative for single LCG candidates leads to overestimated $f_{\rm esc}^{\rm abs}$ values and indicates that any detection of LyC at $z>3.5$ is likely to arise from a more transparent than average sightlines.
  On the other hand, IGM corrections based on maximum transparency to LyC photons give underestimated $f_{\rm esc}^{\rm abs}$ values.
  These results imply that $f_{\rm esc}^{\rm abs}$ measurements for single objects at high redshifts can only be determined with broad range of a values and the only way for improvement is to better understand impact of the IGM on a single line of sight.}
  
  \item{The estimated $f_{\rm esc}^{\rm abs}$ are in the range $\sim5\% - 73\%$ for q1 candidate id 1 and $\sim30\% - 93\%$ for q1 candidate id 368, where different metallicities can produce a variation in $f_{\rm esc}^{\rm abs}$ up to $\sim10\%$.}
  
  \item{Both q1 LCG candidates have EW(Ly$\alpha$) $<$ 50\AA. 
  No clear correlation is seen among EW(Ly$\alpha$), $f_{\rm esc}^{\rm abs}$, and $R_{\rm obs}$ in our relatively small sample. 
  We note that we do not see cases where an LCG candidate has strong EW(Ly$\alpha$) and high $R_{\rm obs}$ ratio.
  With current size of the sample we are unable to exclude the existence of a correlation or anti-correlation between EW(Ly$\alpha$) and LyC leakage into IGM.} 
  
   \item{The stacking procedure of the \textit{Non-detection} candidates did not reveal any significant LyC flux above S/N $>$ 3 in the $u$-band.  
   Based on the results from stacking 39 \textit{Non-detection} candidates with confirmed redshifts, the LyC radiation emitted by galaxies into IGM does not exceed $1\%$ for the average stack.  
   If this is the case for galaxies at $z>6$ then galaxies alone are not able to emit enough LyC radiation to reionize the Universe.}

\end{itemize}

Our analysis demonstrates that the $u-g$ vs $g-i$ colour-colour diagram is useful for identifying $z\sim4$ LCGs.  
However, it emphasizes that detecting clean, reliable sources of LyC radiation is difficult without using high resolution HST imaging in at least one filter, and preferably two or more, as well as spectroscopic redshift confirmation in combination with deep ground or space based LyC photometry. 
Creating larger samples of LCGs at $z>3$ will be possible with large spectroscopic surveys that are followed by high spatial resolution imaging with new ground-based 30m telescopes and space-based telescopes such as the Large UV/Optical/IR Surveyor \citep[LUVOIR;][]{LUVOIR2017} or the Cosmological Advanced Survey Telescope for Optical and UV Research \citep[CASTOR;][]{CASTOR2012}.

\section*{Acknowledgements}
We would like to thank Ikuru Iwata for constructive comments and suggestions that helped to improve this manuscript.
We would like to thank anonymous referee for very useful comments which gave us a possibility to address several issues that were initially overlooked.
AKI is supported by JSPS KAKENHI Grant Number 17H01114.
This research was conducted by the Australian Research Council Centre of Excellence for All Sky Astrophysics in 3 Dimensions (ASTRO 3D), through project number CE170100013.
Support for Program number HST-GO-15100 was provided by NASA through a grant from the Space Telescope Science Institute, which is operated by the Association of Universities for Research in Astronomy, Incorporated, under NASA contract NAS5-26555.
J.C. acknowledges the  Australian Research Council Centre of Excellence for Gravitational Wave Discovery (OzGrav), CE170100004.
The Hyper Suprime-Cam (HSC) collaboration includes the astronomical communities of Japan and Taiwan, and Princeton University. The HSC instrumentation and software were developed by the National Astronomical Observatory of Japan (NAOJ), the Kavli Institute for the Physics and Mathematics of the Universe (Kavli IPMU), the University of Tokyo, the High Energy Accelerator Research Organization (KEK), the Academia Sinica Institute for Astronomy and Astrophysics in Taiwan (ASIAA), and Princeton University. Funding was contributed by the FIRST program from Japanese Cabinet Office, the Ministry of Education, Culture, Sports, Science and Technology (MEXT), the Japan Society for the Promotion of Science (JSPS), Japan Science and Technology Agency (JST), the Toray Science Foundation, NAOJ, Kavli IPMU, KEK, ASIAA, and Princeton University.
This work is based on observations obtained with MegaPrime/MegaCam, a joint project of CFHT and CEA/DAPNIA, at the Canada-France-Hawaii Telescope (CFHT) which is operated by the National Research Council (NRC) of Canada, the Institut National des Sciences de l'Univers of the Centre National de la Recherche Scientifique (CNRS) of France, and the University of Hawaii. This work is based in part on data products produced at TERAPIX and the Canadian Astronomy Data Centre as part of the Canada-France-Hawaii Telescope Legacy Survey, a collaborative project of NRC and CNRS.
This paper makes use of software developed for the Large Synoptic Survey Telescope.  We thank the LSST Project for making their code available as free software at  http://dm.lsst.org
The Pan-STARRS1 Surveys (PS1) have been made possible through contributions of the Institute for Astronomy, the University of Hawaii, the Pan-STARRS Project Office, the Max-Planck Society and its participating institutes, the Max Planck Institute for Astronomy, Heidelberg and the Max Planck Institute for Extraterrestrial Physics, Garching, The Johns Hopkins University, Durham University, the University of Edinburgh, Queen’s University Belfast, the Harvard-Smithsonian Center for Astrophysics, the Las Cumbres Observatory Global Telescope Network Incorporated, the National Central University of Taiwan, the Space Telescope Science Institute, the National Aeronautics and Space Administration under Grant No. NNX08AR22G issued through the Planetary Science Division of the NASA Science Mission Directorate, the National Science Foundation under Grant No. AST-1238877, the University of Maryland, and Eotvos Lorand University (ELTE) and the Los Alamos National Laboratory.
Based [in part] on data collected at the Subaru Telescope and retrieved from the HSC data archive system, which is operated by Subaru Telescope and Astronomy Data Center at National Astronomical Observatory of Japan.
Based on data obtained with the European Southern Observatory Very Large Telescope, Paranal, Chile, under Large Program 185.A-0791, and made available by the VUDS team at the CESAM data center, Laboratoire d'Astrophysique de Marseille, France.
The HST data matched to the VUDS-DR1 are described in Grogin et al. (2011) and Koekemoer et al. (2011) for CANDELS and include data from the ERS (Windhorst et al. 2011).
This work is based on observations taken by the 3D-HST Treasury Program (HST-GO-12177 and HST-GO-12328) with the NASA/ESA Hubble Space Telescope, which is operated by the Association of Universities for Research in Astronomy, Inc., under NASA contract NAS5-26555.




\bibliographystyle{mnras}
\bibliography{example.bib} 





\begin{appendix}
\section{Images of the q1 and q2 candidates in other available filters}

\begin{figure*}
\begin{center}
    \subfloat{\includegraphics[width=.25\textwidth]{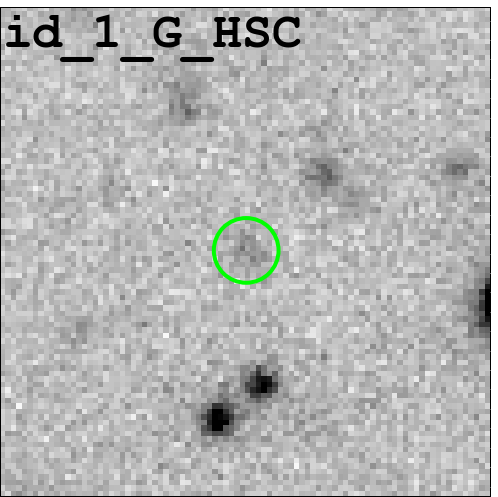}}
    \hspace{-0.05cm}\vspace{-0.05cm}
    \subfloat{\includegraphics[width=.25\textwidth]{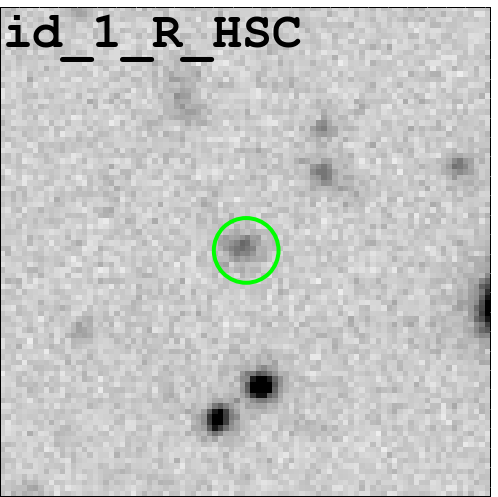}}
    \hspace{-0.05cm}\vspace{-0.05cm}
    \subfloat{\includegraphics[width=.25\textwidth]{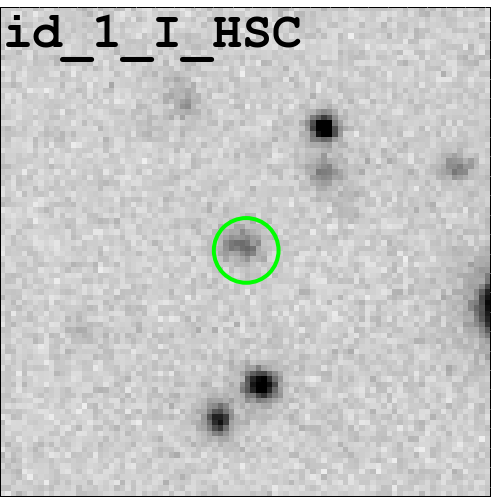}}
    \hspace{-0.05cm}\vspace{-0.05cm}
    \subfloat{\includegraphics[width=.25\textwidth]{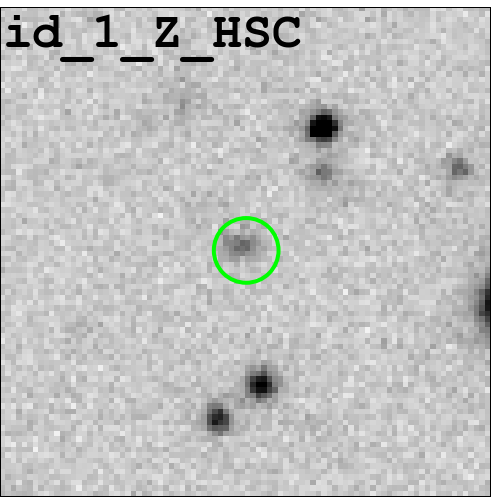}}
    \hspace{-0.05cm}\vspace{-0.05cm}
    
    \subfloat{\includegraphics[width=.25\textwidth]{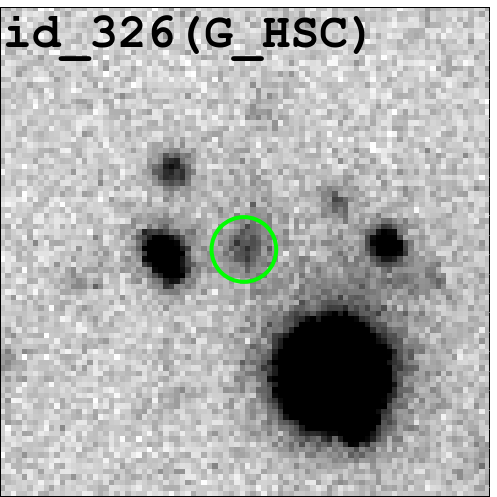}}
    \hspace{-0.05cm}\vspace{-0.05cm}
    \subfloat{\includegraphics[width=.25\textwidth]{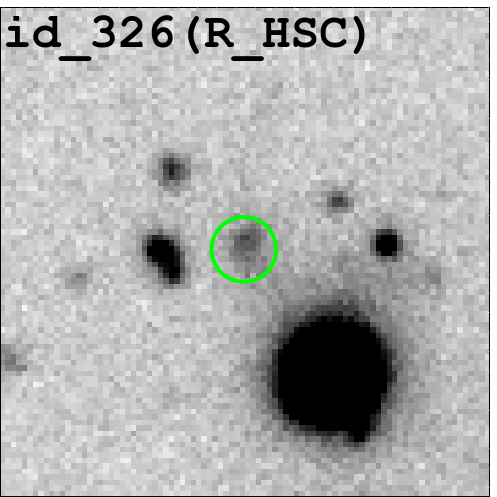}}
    \hspace{-0.05cm}\vspace{-0.05cm}
    \subfloat{\includegraphics[width=.25\textwidth]{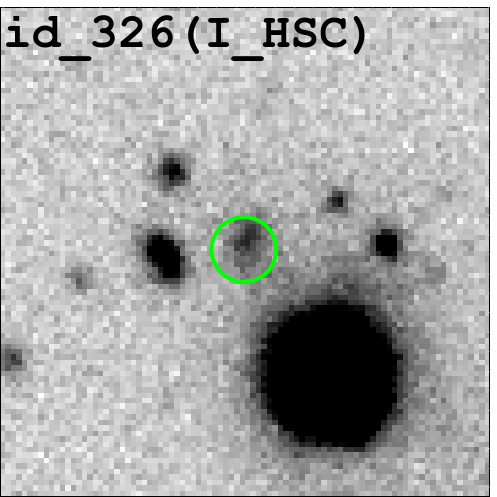}}
    \hspace{-0.05cm}\vspace{-0.05cm}
    \subfloat{\includegraphics[width=.25\textwidth]{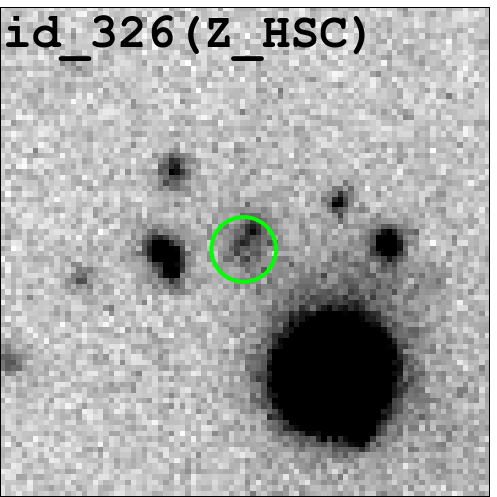}}
    \hspace{-0.05cm}\vspace{-0.05cm}
    
    \subfloat{\includegraphics[width=.25\textwidth]{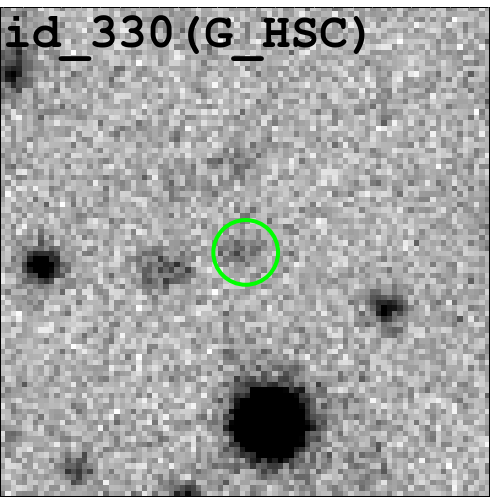}}
    \hspace{-0.05cm}\vspace{-0.05cm}
    \subfloat{\includegraphics[width=.25\textwidth]{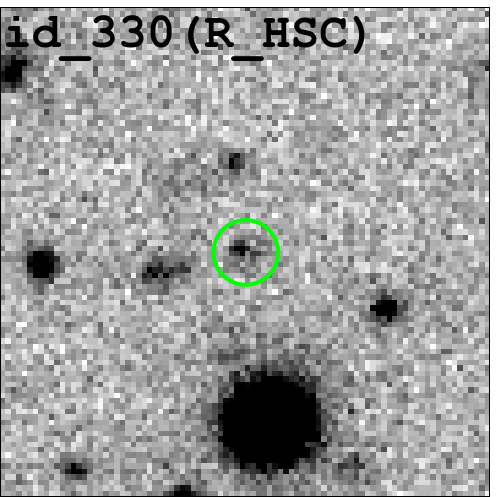}}
    \hspace{-0.05cm}\vspace{-0.05cm}
    \subfloat{\includegraphics[width=.25\textwidth]{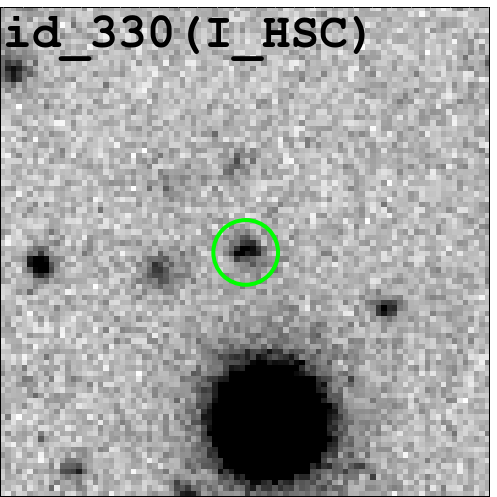}}
    \hspace{-0.05cm}\vspace{-0.05cm}
    \subfloat{\includegraphics[width=.25\textwidth]{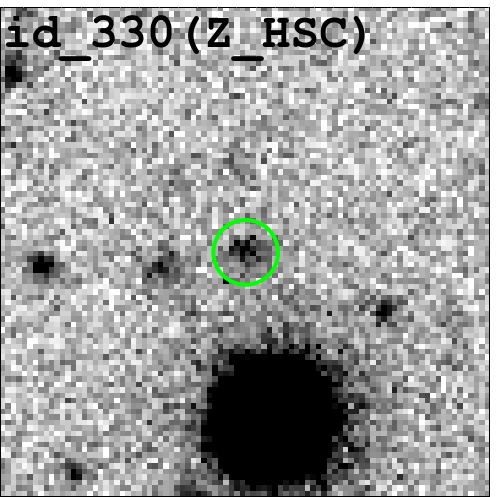}}
    \hspace{-0.05cm}\vspace{-0.05cm}

    \subfloat{\includegraphics[width=.25\textwidth]{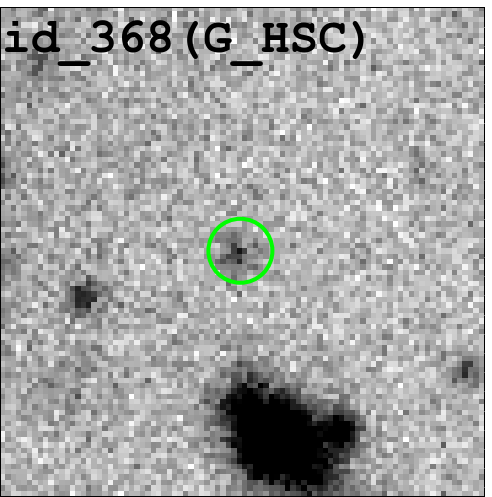}}
    \hspace{-0.05cm}\vspace{-0.05cm}
    \subfloat{\includegraphics[width=.25\textwidth]{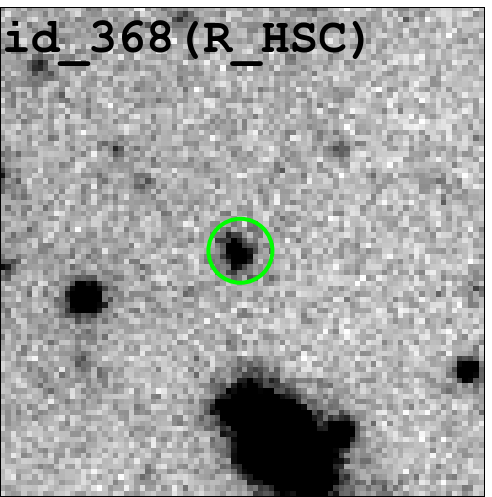}}
    \hspace{-0.05cm}\vspace{-0.05cm}
    \subfloat{\includegraphics[width=.25\textwidth]{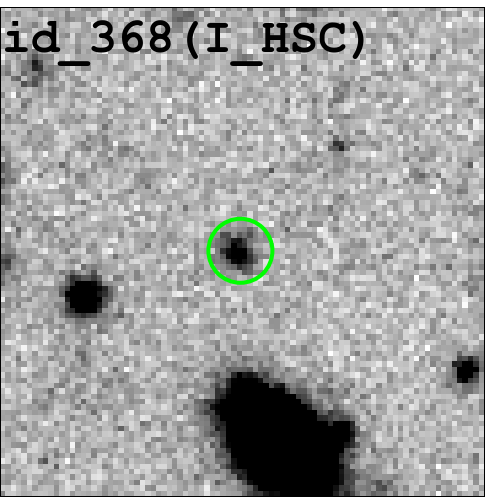}}
    \hspace{-0.05cm}\vspace{-0.05cm}
    \subfloat{\includegraphics[width=.25\textwidth]{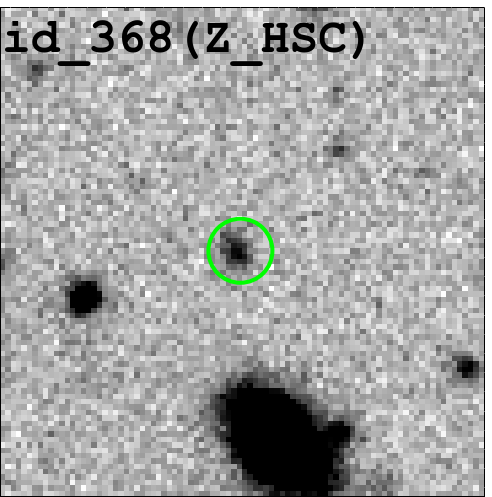}}
    \hspace{-0.05cm}\vspace{-0.05cm}

    \subfloat{\includegraphics[width=.25\textwidth]{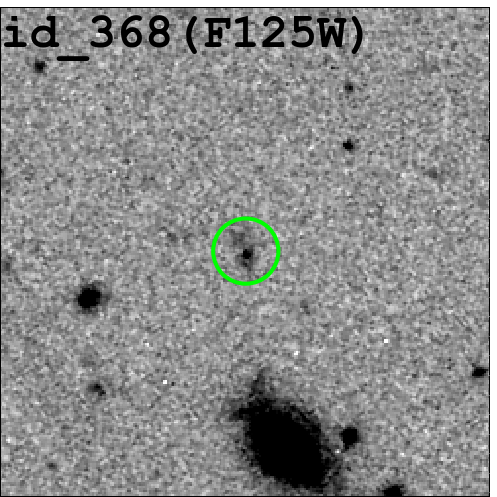}}
    \hspace{-0.05cm}\vspace{-0.05cm}
    \subfloat{\includegraphics[width=.25\textwidth]{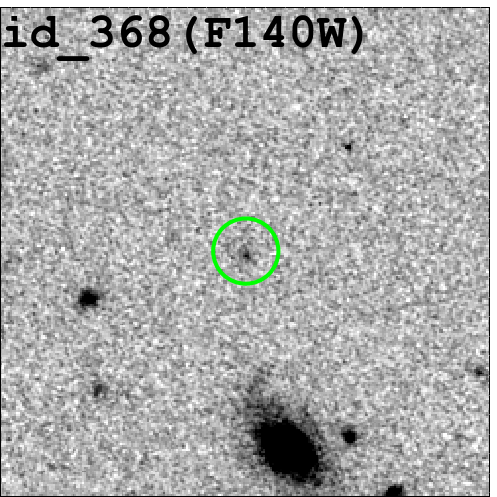}}
    \hspace{-0.05cm}\vspace{-0.05cm}
    \subfloat{\includegraphics[width=.25\textwidth]{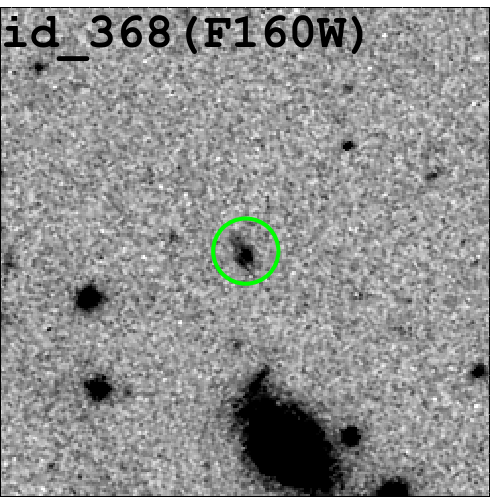}}
    \hspace{-0.05cm}\vspace{-0.05cm}
    \subfloat{\includegraphics[width=.25\textwidth]{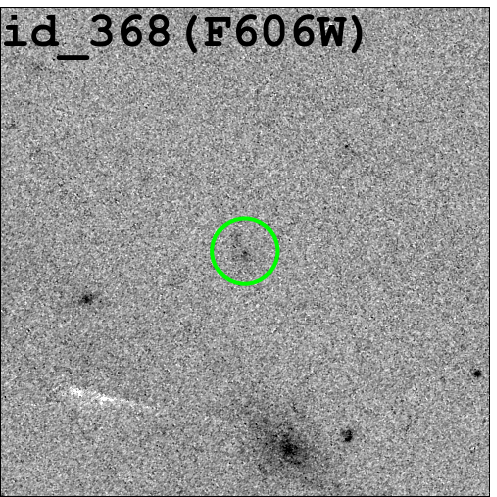}}
    \hspace{-0.05cm}\vspace{-0.05cm}

\caption{Cutouts of the q1 (id 368, id 1) and q2 (id 326, id 330, id 421) candidates in $g$-HSC, $r$-HSC, $i$-HSC and $z$-HSC as well HST F125W, F140W, 160W and  F606W where available. The green circles are $2^{\prime\prime}$ in diameter and thumbnails are $15^{\prime\prime}\times15^{\prime\prime}$ in size.}
\label{A1}
\end{center}
\end{figure*}

\begin{figure*}
\begin{center}

    \subfloat{\includegraphics[width=.25\textwidth]{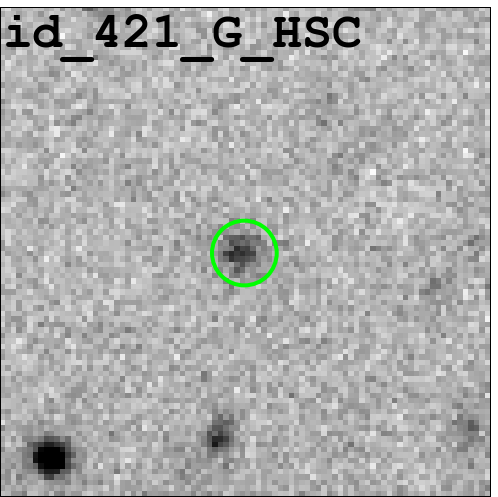}}
    \hspace{-0.05cm}\vspace{-0.05cm}
    \subfloat{\includegraphics[width=.25\textwidth]{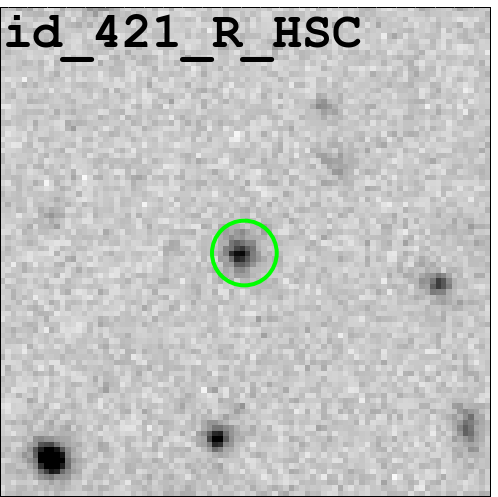}}
    \hspace{-0.05cm}\vspace{-0.05cm}
    \subfloat{\includegraphics[width=.25\textwidth]{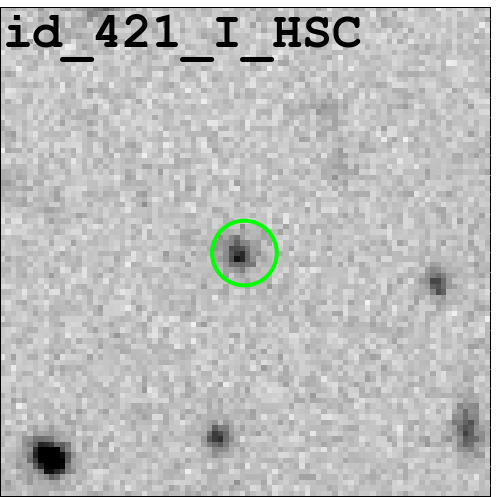}}
    \hspace{-0.05cm}\vspace{-0.05cm}
    \subfloat{\includegraphics[width=.25\textwidth]{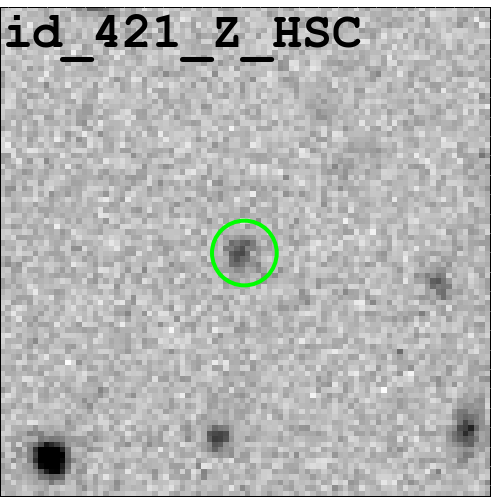}}
    \hspace{-0.05cm}\vspace{-0.05cm}

\caption{Same as \ref{A1}.}
\label{figA2}
\end{center}
\end{figure*}

\section{Spectra of the q1 and q2 candidates}

\begin{figure*}
\begin{center}

    \subfloat{\includegraphics[ width=18cm]{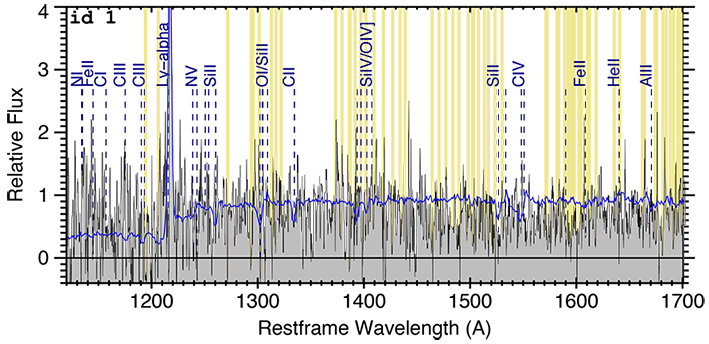}}
    \vspace{1cm}
    \subfloat{\includegraphics[ width=18cm]{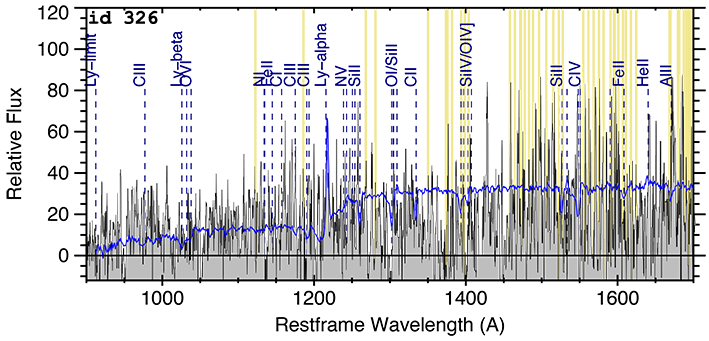}}

\caption{1D spectrum of the selected 5 LCG candidates (q1 and q2). Composite spectrum (blue) is fitted to the candidate spectrum (grey) to confirm reported redshift, detect Ly$\alpha$ forest or other spectroscopic features (blue dashed lines). With yellow vertical stripes positions of the sky lines are marked.}
\label{B1}
\end{center}
\end{figure*}

\newpage

\begin{figure*}
\begin{center}

    \subfloat{\includegraphics[ width=18cm]{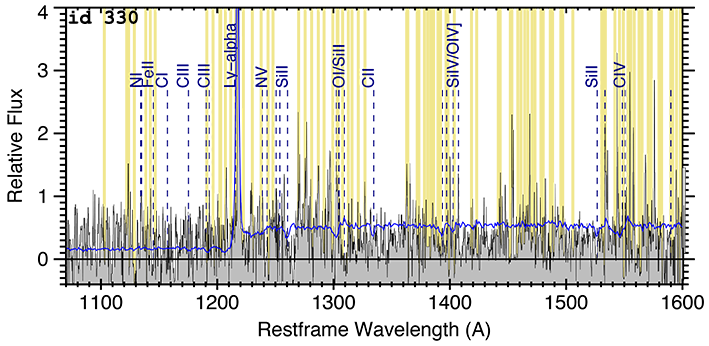}}
    \vspace{1cm}
    \subfloat{\includegraphics[ width=17.5cm]{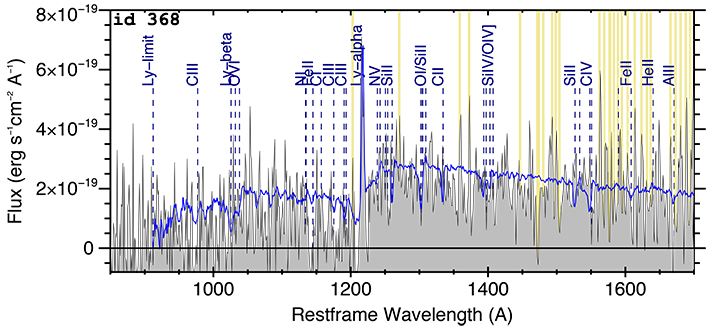}}

\caption{Same as \ref{B1}.}
\label{B5}
\end{center}
\end{figure*}

\newpage

\begin{figure*}
\begin{center}

    \vspace{-0.4cm}
    \subfloat{\includegraphics[ width=18cm]{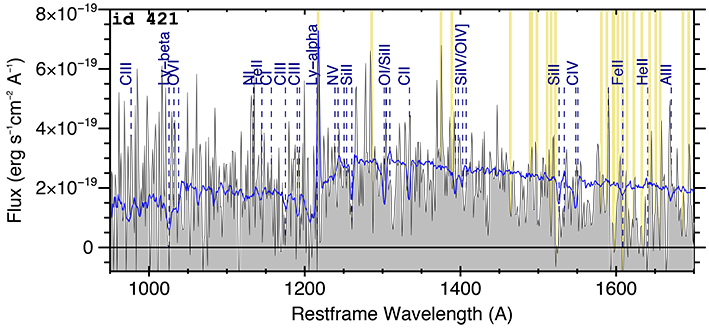}}

\caption{Same as \ref{B1}.}
\label{B6}
\end{center}
\end{figure*}

\begin{figure*}
\begin{center}

    \vspace{-0.4cm}
    \subfloat{\includegraphics[ width=18cm]{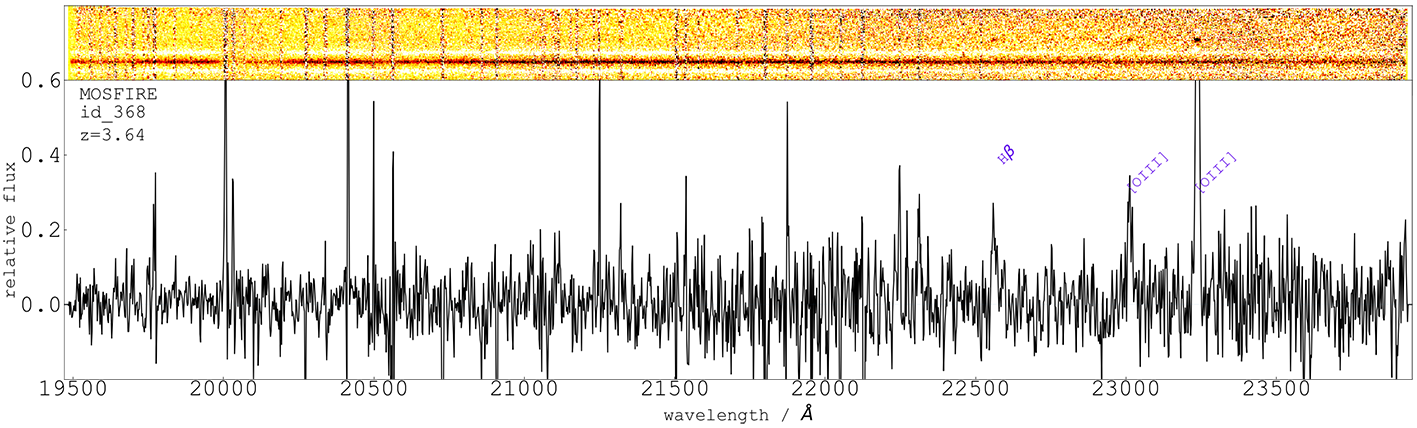}}

\caption{Keck MOSFIRE 1D and 2D spectrum of the candidate id 368 used to additionally confirm redshift of the candidate. Positions of the H$\beta$ and [OIII] lines are marked. }
\label{B6}
\end{center}
\end{figure*}

\end{appendix}

\bsp	
\label{lastpage}
\end{document}